\documentclass[english,structabstract]{aa}
\usepackage{mathptmx}
\usepackage[T1]{fontenc}
\usepackage[utf8]{inputenc}
\usepackage{amssymb}
\usepackage{graphicx}

\makeatletter
\bibpunct{(}{)}{;}{a}{}{,} % to follow the A&A style
\newcommand{\teff}{T_{\mathrm{eff}}}
\newcommand{\logg}{\log g}
\newcommand{\feh}{\left[\mathrm{Fe}/\mathrm{H}\right]}
\newcommand{\ltaur}{\log\tau_{\mathrm{Ross}}}
\newcommand{\taur}{\tau_{\mathrm{Ross}}}
\newcommand{\kapr}{\kappa_{\mathrm{Ross}}}

\newcommand{\vzrms}{v_{z,\mathrm{rms}}}
\newcommand{\vzrmsp}{v_{z,\mathrm{rms}}^{\mathrm{peak}}}
\newcommand{\nab}{\vec{\nabla}}
\newcommand{\nsad}{\vec{\nabla}_{\mathrm{sad}}}

\newcommand{\nsadp}{\vec{\nabla}_{\mathrm{sad}}^{\mathrm{peak}}}
\newcommand{\hav}{\left\langle \mathrm{3D}\right\rangle}
\newcommand{\havz}{\hav_z}
\newcommand{\havr}{\hav_{\mathrm{Ross}}}
\newcommand{\havf}{\hav_{\mathrm{500}}}
\newcommand{\havcm}{\hav_{m}}
\newcommand{\havhse}{\hav_{\mathrm{HSE}}}
\newcommand{\nltau}{\log\tilde{\tau}}
\newcommand{\angs}{\mathring{A}}

\newcommand{\rhoc}{\delta \rho_{\mathrm{rms}}}

\newcommand{\eqw}{W_\lambda}
\newcommand{\lgf}{\log gf}
\newcommand{\dlga}{\Delta\log\varepsilon}
\newcommand{\fei}{\ion{Fe}{i}}
\newcommand{\feii}{\ion{Fe}{ii}}
\newcommand{\mturb}{\xi_{\mathrm{turb}}}

\newcommand{\xex}{\chi_{\mathrm{exc}}}

\usepackage{hyperref}
\usepackage{txfonts}

\makeatother

\usepackage{babel}
\begin{document}

\title{The \textsc{Stagger}-grid: A grid of 3D stellar atmosphere models}

\subtitle{II. Horizontal and temporal averaging and spectral line formation}

\author{Z. Magic\inst{1,2}, R. Collet\inst{2,1}, W. Hayek\inst{1,2} \and 
M. Asplund\inst{2,1}}

\institute{Max-Planck-Institut für Astrophysik, Karl-Schwarzschild-Str. 1, 85741
Garching, Germany\\
\email{magic@mpa-garching.mpg.de} \and  Research School of Astronomy
\& Astrophysics, Cotter Road, Weston ACT 2611, Australia}

\offprints{magic@mpa-garching.mpg.de}

\date{Received ...; Accepted...}

\abstract{}{We study the implications of averaging methods with different
reference depth scales for 3D hydrodynamical model atmospheres computed
with the \textsc{Stagger}-code. The temporally and spatially averaged
(hereafter denoted as $\hav$) models are explored in the light of
local thermodynamic equilibrium (LTE) spectral line formation by comparing
spectrum calculations using full 3D atmosphere structures with those
from $\hav$ averages. }{We explored methods for computing mean
$\hav$ stratifications from the \textsc{Stagger}-grid time-dependent
3D radiative hydrodynamical atmosphere models by considering four
different reference depth scales (geometrical depth, column-mass density,
and two optical depth scales). Furthermore, we investigated the influence
of alternative averages (logarithmic, enforced hydrostatic equilibrium,
flux-weighted temperatures). For the line formation we computed curves
of growth for $\fei$ and $\feii$ lines in LTE . }{The resulting
$\hav$ stratifications for the four reference depth scales can be
very different. We typically find that in the upper atmosphere and
in the superadiabatic region just below the optical surface, where
the temperature and density fluctuations are highest, the differences
become considerable and increase for higher $\teff$, lower $\logg$,
and lower $\feh$. The differential comparison of spectral line formation
shows distinctive differences depending on which $\hav$ model is
applied. The averages over layers of constant column-mass density
yield the best mean $\hav$ representation of the full 3D models for
LTE line formation, while the averages on layers at constant geometrical
height are the least appropriate. Unexpectedly, the usually preferred
averages over layers of constant optical depth are prone to increasing
interference by reversed granulation towards higher effective temperature,
in particular at low metallicity.}{}


\maketitle

\section{Introduction\label{sec:Introduction}}

Theoretical model atmospheres are needed in order to interpret stellar
fluxes and derive individual characteristics of stars, like stellar
parameters and chemical abundances. In recent decades, successive
improvements of the often used one-dimensional (1D) hydrostatic atmosphere
models have confirmed their predictive capabilities \citep[see, e.g., ][]{Gustafsson:2008p3814}
but also highlighted their limitations. In fact, these 1D models make
use of several simplifications in favor of computational ease, the
most prominent one being the treatment of convection with the mixing-length
theory \citep[MLT,][]{BohmVitense:1958p4822,Henyey:1965p15592}. The
latter entails several free parameters, in particular the free mixing-length
parameter, $\alpha_{\mathrm{MLT}}$, which is a priori unknown, hence
normally calibrated for the Sun by observations and assumed constant
for all stars. Moreover, the calculation of synthetic spectral absorption
lines in 1D requires the additional calibration of micro- and macro-turbulence
parameters ($\xi_{\mathrm{turb}}$ and $\chi_{\mathrm{turb}}$, respectively)
in order to properly account for the contribution of non-thermal convective
and turbulent motions to the broadening of spectral line profiles.

Most of the limitations of 1D modeling of convection can be overcome
only by performing time-dependent, three dimensional (3D), radiative-hydrodynamical
(RHD) calculations \citep[see][and references therein]{Nordlund:2009p4109}.
The goal of 3D simulations is to provide realistic ab initio models
where stellar surface convection emerges self-consistently from first
principles. Compared to 1D models, such 3D RHD models are able, for
the Sun in particular, to predict additional observable features of
stars associated with stellar surface velocity fields and temperature
and density inhomogeneities, e.g. surface granulation pattern, line
asymmetries, and center-to-limb variation \citep[CLV; e.g., such as][]{Asplund:2000p20866,Pereira:2013arXiv1304}.
To systematically study such properties of stars with a realistic
approach, we computed a large grid of 3D models using the \textsc{Stagger}-code,
covering a wide range in stellar parameters%
\footnote{In the following, we always refer to stellar \emph{atmospheric} parameters.%
} ($\teff$, $\logg$, and $\feh$) for late-type (spectral type FGK)
stars \citep[see][hereafter Paper I]{Magic:2013}.

It is advantageous to reduce the relatively large amount of data from
the full 3D atmospheric models to temporally and spatially averaged
(hereafter $\hav$) representations. However, this reduction comes
at the expense of physical self-consistency \citep[see][]{Atroshchenko:1994p14010}.
Nonetheless, in this way one can deal with more manageable atmospheric
data structures compared to the otherwise enormous amount of information
associated with the full 3D models. These mean $\hav$ stratifications
are usually compared with classical 1D hydrostatic atmosphere models.
\citet{Nordlund:2001p6371} point out that the large-amplitude fluctuations
in the superadiabatic region%
\footnote{The SAR can be located with the superadiabatic gradient, e.g., with
$\nsad>0.1\max\left[\nsad\right]$ one obtains typically a range of
$-0.5\lesssim\ltaur\lesssim4.0$.%
} (SAR) leads to deviations from the hydrostatic equilibrium. Furthermore,
the 3D data sets incorporate quantities emerging from the hydrodynamics
and associated with convection itself, such as, self-consistent velocity
fields and turbulent pressure, for which there are no physically consistent
counterparts in the case of 1D hydrostatic models.

The definition of the $\hav$ stratifications is neither unambiguous
nor unique, but depends largely on the choice of reference depth scale.
When dealing with the analysis of the atmospheric layers above the
optical surface, monochromatic or Rosseland optical depth scales are
usually considered the appropriate choice since these are the natural
reference depth scales that are used to describe radiative transfer
processes in the photosphere. On the other hand, the optical depth
loses its usefulness somewhat in the very deep optically thick layers
below the optical surface, since here the mean free path of photons
becomes very short and the radiative transfer insignificant. Therefore,
other reference scales are best suited to describing the main properties
of the stellar stratification. Also, the bimodal and highly asymmetric
distribution of bulk upflows and of downflows in the convective zone
complicates the definition of a meaningful unique average value, particularly
near the surface, at the transition between convectively unstable
and stable regions.

\citet{Uitenbroek:2011p10448} investigated the application of $\left\langle 3\mathrm{D}\right\rangle $
models to spectral line formation. They computed and compared continuum
and atomic line intensities and their respective CLV from $\left\langle 3\mathrm{D}\right\rangle $
and 3D models. They conclude that a mean $\hav$ stratification is
insufficient to represent the full 3D atmosphere model in the light
of spectral analysis. As reasons for the latter they list the non-linearity
of the Planck function, formation of molecules, and the asymmetry
of convective motions.

The present work constitutes the second paper in the \textsc{Stagger}-grid
series. Here, we want to explore the following key question: which
averaging method leads to the closest $\left\langle 3\mathrm{D}\right\rangle $
representation of a full 3D data set in the light of spectral line
formation calculations? Therefore, we investigate spectral line absorption
features by probing the latter with fictitious $\fei$ and $\feii$
lines with different strengths and excitation potentials for different
stellar parameters.

\section{Averaging 3D models\label{sec:Methods}}

\label{sub:stagger-code}The 3D models that form the basis of the
present work were computed with the \textsc{Stagger-}code. For a general
description of our grid of 3D models, we refer the reader to Paper
I. In short, the \textsc{Stagger-}code solves the time-dependent,
3D hydrodynamical equations coupled with realistic non-gray radiative
transfer. We utilize an updated version of the realistic state-of-the-art
equation of state (EOS) by \citet{Mihalas:1988p20892}. Continuum
and sampled line opacity are taken primarily from the MARCS package
\citep[see also references in Paper I]{Gustafsson:2008p3814}. The
radiative transfer is solved for nine angles along long characteristics
with a slightly modified version of the \citet{Feautrier:1964p21596}
method. The opacity-binning method with 12 opacity bins is applied
to all \textsc{Stagger}-grid models to reduce the computational burden
while still accounting for non-gray radiative transfer under the assumption
of local thermodynamic equilibrium (LTE); in particular, the effects
of scattering are neglected \citep[see][]{Nordlund:1982p6697,Skartlien:2000p9857,Collet:2011p6147}.
Our simulations are of the so-called \textit{box-in-a-star} type,
and they cover only a small representative volume of stellar surface
that typically includes about ten granules horizontally and spans
about 14 pressure scale heights vertically. The numerical resolution
of the Cartesian grid is $240^{3}$. It features a non-equidistant
vertical axis in order to enhance resolution in the layers with the
steepest temperature gradients. The vertical boundaries are open,
while the horizontal ones are periodic.

\subsection{Computing temporal and horizontal averages\label{sub:aAveraging}}

We computed various temporal and horizontal averages for a large number
of physical quantities of interest. For the spatial (horizontal) averages,
we computed $\hav$ stratifications by considering four different
reference depth scales and averaging the various physical quantities
on layers of constant
\begin{itemize}
\item geometrical height, $z$;
\item column mass density, $m=\int\rho\, dz$;
\item Rosseland optical depth, $\taur=\int(\rho\kappa_{\mathrm{Ross}})\, dz$;
\item optical depth at 500 nm, $\tau_{500}=\int(\rho\kappa_{500})\, dz$,
\end{itemize}
(hereafter denoted by $\havz$, $\havcm$, $\havr$, and $\havf$,
respectively), where $\rho$ is the gas density, and $\kappa_{\mathrm{Ross}}$
and $\kappa_{500}$ are the Rosseland mean opacity%
\footnote{Including both line and continuum opacity.%
} and opacity at 500 nm, respectively, both defined as cross-sections
per unit mass. 

The geometrical averages $\left\langle 3\mathrm{D}\right\rangle _{z}$
are easily taken directly from the output of the \textsc{Stagger}-code,
since the numerical mesh of this code is Eulerian in nature. For the
three other (Lagrangian-like) averages, the original data sets have
to be remapped to their respective new reference depth scale by individually
interpolating each column of each 3D simulation snapshot (see \ref{sub:Interpolation-new-reference-scale}).
Furthermore, we also considered four additional averages:
\begin{itemize}
\item flux-weighted average temperature, $\langle T^{4}\rangle$;
\item average brightness temperature at 500nm, $\langle T_{\mathrm{rad}}\rangle$;
\item logarithmic average, $\hav_{\log}$; and
\item enforced-hydrostatic-equilibrium average, $\havhse$. 
\end{itemize}
We determine the flux-weighted temperature stratification $\langle T^{4}\rangle$
by evaluating the spatial averages of $T^{4}$, motivated by the Stefan-Boltzmann
law for wavelength-integrated radiative flux. The brightness temperature
average $T_{\mathrm{rad}}$ is computed using the expression $B_{500}^{-1}\left(\left\langle B_{500}(T)\right\rangle \right)$,
where $B_{500}$ and $B_{500}^{-1}$ denote the Planck function at
500 nm and its inverse, respectively (see also Sect. \ref{sub:Temperature}).
The depth-dependent $\langle T_{\mathrm{rad}}\rangle$ thus needs
to be interpreted as the equivalent brightness temperature corresponding
to the average black-body emission at 500 nm from each layer. For
$\hav_{\log}$ we define spatial averages of a given 3D variable $X$
as $\exp\left(\left\langle \log{X}\right\rangle \right)$. Finally,
since the $\hav$ models do not in general fulfill the hydrostatic
equilibrium condition (see App. \ref{app:hse_stratification}), for
the $\havhse$ averages we \emph{enforce} hydrostatic equilibrium
by adjusting the density and adjusting the thermodynamic pressure
$p_{\mathrm{th}}$ consistently with the EOS, until hydrostatic equilibrium
is attained. We emphasize that the proper enforcement of hydrostatic
equilibrium requires that one considers both the thermodynamic $p_{\mathrm{th}}$
and turbulent $p_{\mathrm{turb}}$ contributions to total pressure
$p_{\mathrm{tot}}$: the gas pressure in the atmosphere is in fact
significantly reduced because of the structural support provided by
turbulent pressure. Then, a new geometrical depth $z$ is computed
(see Eq. \ref{eq:hse}).

Classical hydrostatic 1D models of stellar atmospheres are often defined
and computed on an optical depth scale, since this allows the numerical
resolution to be easily adjusted where it is most needed to achieve
the highest accuracy in the solution of the radiative transfer equation
in the atmospheric layers, both during the modeling itself and during
line-formation calculations. Therefore, especially for radiative transfer-oriented
applications, these 1D models can be compared most naturally with
averages of corresponding 3D models on constant optical depth, $\havr$
or $\havf$. In Paper I, in particular, we adopted $\havr$ as our
standard averaging choice. One of the main reasons we chose $\havr$
over $\havf$ is that during the scaling of the simulations and the
construction of the initial snapshots, the top physical boundary of
essentially all models reached up to $\left\langle \ltaur\right\rangle _{\mathrm{top}}\approx-6.0$
(see Paper I). In contrast, the vertical extent of the simulations
in terms of optical depth at $500$~nm varies depending on stellar
parameters ($\logg$ in particular) owing to the concomitant variations
in opacity at $500$~nm as a function of temperature and density.
Therefore, the $\havf$ models in general require a careful extrapolation
at the top to be extended up to $\log\tau_{500}{\approx}-6.0$ (see
Sect. \ref{sub:Extrapolation-at-the-top}).

While $\havr$ or $\havf$ represent natural reference depth scales
for the mean photospheric stratification, $\havz$ or $\havcm$ is
better suited to describing the average physical conditions below
the stellar surface; e.g., only the geometrical averages fulfill conservation
of momentum and energy (see App. \ref{app:hse_stratification}).

In late-type stellar atmospheres, the continuum opacity $\kappa_{\lambda}$
in the optical is dominated by the $\mathrm{H}^{-}$ bound-free absorption
that is sensitive to temperature ($\sim T^{10}$). Therefore, even
small fluctuations in $T$ will result in large variations in $\kappa_{\lambda}$,
which in turn will lead to a high degree of spatial corrugation of
layers at constant optical depth \citep[see][]{Stein:1998p3801}.
Furthermore, owing to such highly non-linear behavior of the $\mathrm{H}^{-}$
opacity, temperature fluctuations around the average will be reduced
by interpolation to layers of constant optical depth (see Sect. \ref{sub:Contrast}).

We note briefly that only the geometrical averages $\havz$, sampled
over a sufficient time length, preserve the conservation properties
of the hydrodynamical equations, such as hydrostatic equilibrium and
conservation of energy. Furthermore, depending on the intended particular
application of $\hav$ models, it is very important to use these carefully,
since the different types of $\hav$ models vary significantly among
the different averaging methods.

\subsection{Basic averaging procedure\label{sub:Averaging-procedure}}

We proceeded with the following steps in order to obtain the $\hav$
models: 
\begin{enumerate}
\item Retrieval of 3D variables of interest; 
\item Interpolation to new reference depth scale; 
\item Computation of horizontal averages and statistics; 
\item Extrapolation of horizontal averages, if necessary; 
\item Computation of temporal averages. 
\end{enumerate}
In case of the geometrical averages $\havz$, steps 2 and 4 are unnecessary
and are therefore skipped. Owing to the generally non-linear response
of the various physical quantities as a function of basic independent
variables and the EOS, the interpolation to a new reference depth
scale should be performed after retrieving the variables. In particular,
because of these non-linearities, we caution against the derivation
of thermodynamic variables via the EOS by utilizing averaged independent
variables interpolated to the new reference depth scale, since the
spatial averaging will inevitably break the physical self-consistency
present in the full original 3D data (see Sect. \ref{sub:Interpolation-new-reference-scale}
and Appendix \ref{app:Deviations-from-EOS}).

At the vertical boundaries of our simulation box are so-called \textit{ghost
zones}, each consisting of five layers at the top and bottom. Their
sole purpose is to numerically define the boundary conditions at both
vertical ends. They do not contain physically meaningful values, so
we excluded them before the averaging procedure.

To speed up the calculations without noticeably degrading the statistical
properties, when computing the averages we considered only every fourth
column of the 3D data cubes in both horizontal directions ($x$ and
$y$), which means that the initial $N_{x}N_{y}=240^{2}$ columns
are reduced down to $60^{2}$. The vertical extent of the columns
is unchanged with $N_{z}=230$ (geometrical) or $101$ (all other
reference depth scales). Tests ensured that this horizontal reduction
does not influence the horizontal averages owing to the still large
sample of vertical columns considered and the multiple snapshots included
in the temporal averaging.

For step 3, we used an arithmetic mean to compute the average values
of variable $X$ for snapshot $t$ at each horizontal layer $z$:
\begin{equation}
\left\langle X\right\rangle _{z,t}=\frac{1}{N_{x}N_{y}}\sum_{x=1}^{N_{x}}\sum_{y=1}^{N_{y}}X_{xyz,t}\label{eq:spatial}
\end{equation}
with $N_{x}$ and $N_{y}$ the number of horizontal elements. For
exponentially varying variables like density and pressure, we computed
also logarithmic averages, i.e., replacing $X_{xyz}$ with $\log X_{xyz}$
in Eq. \ref{eq:spatial}, denoting the models with $\hav_{\log}$.
In the final step 5, temporal averages are evaluated with 
\begin{equation}
\left\langle X\right\rangle _{z}=\frac{1}{N_{t}}\sum_{t=1}^{N_{t}}\left\langle X\right\rangle _{z,t}\label{eq:temporal}
\end{equation}
with $N_{t}\approx100-150$ being the total number of snapshots considered
for each simulation, which corresponds typically to about two turnover
times. In the present work, the combined temporal and spatial averages
of variable $X$ are always denoted with $\left\langle X\right\rangle _{\tilde{z}}$,
where $\tilde{z}$ is the considered reference depth scale.

Since the 3D structures display a great plethora of details, for each
relevant 3D variable we also determine a number of additional statistical
properties (standard deviation $\sigma$, root mean square, minimum-maximum
range, and histograms of the distribution of values) at each horizontal
layer, which are presented and discussed in Sect. \ref{sec:Statistical-properties}.
As for the spatial averages, the standard deviation and the root mean
square are evaluated in step $3$ for each layer $z$ using the same
basic expression as in Eq.~\ref{eq:spatial} and, if necessary, doubly
extrapolated at the top as in steps 2 and 4 (see Sect. \ref{sub:Extrapolation-at-the-top}).
Finally, their temporal averages are computed in step 5.

Histograms of the distribution of values we determined separately,
and we use temporal averages of the depth-dependent extrema of variable
$X$, $\left\langle \min X\right\rangle _{z}$ and $\left\langle \max X\right\rangle _{z}$
to define a depth-dependent range $r_{z}=\left[\left\langle \min X\right\rangle _{z},\left\langle \max X\right\rangle _{z}\right]$
for the histograms. For the 3D variable $X$ at time $t$, we determined
a set of 1D histograms, $p_{r,z,t}\left(X\right)$, for each individual
layer $z$. The depth-dependent range $r_{z}$ is resolved with $N_{r}=200$
equidistant points; temporal averages $p_{r,z}\left(X\right)$ of
the histograms are computed using a subset of $N_{t}=20$ equidistant
snapshots (see Sect. \ref{sub:Histograms} for details). 

Finally, we also computed averages and associated statistical properties
separately for up- and downflows, which we differentiate based on
the sign of the vertical component of the velocity. Of course, when
computing such averages and statistics, one has to account for the
correct filling factor in either case, i.e. for the number of elements
$N_{x,y}$ belonging to up- or downflows, respectively (Sect. \ref{sub:Up-and-downflows}).

\subsection{Interpolation to the new reference depth scale\label{sub:Interpolation-new-reference-scale}}

To interpolate to the new reference depth scale (hereafter denoted
as $\tilde{z}$) in step 2, we defined a new equidistant logarithmic
reference optical depth scale, $\tilde{z}=\nltau$, from $-5.0,\dots,+5.0$
in steps of $0.1$ for both optical depth scales $\taur$ and $\tau_{500}$.
In the case of averaging based on the column-mass density scale $m$,
we used the column-mass density $\tilde{m}$ normalized to the mean
value of $m$ at the optical surface, i.e. $\tilde{z}=\log(\tilde{m})=\log(m/\left\langle m\right\rangle _{\mathrm{surf}})$
for the new reference depth scale, where $\left\langle m\right\rangle _{\mathrm{surf}}$
was determined at $\left\langle \tau_{\mathrm{Ross}}=0\right\rangle $
and considered a fixed range from $-3.0,\dots,+2.0$ in steps of 0.05
for all simulations. All variables, $X$, we remapped column-wise
from the original geometrical depth scale to the new reference depth
scale, namely $X_{xy}\left(z\right)\rightarrow\tilde{X}_{xy}\left(\tilde{z}\right)$.
We use linear interpolation, since quadratic interpolation introduced
numerical artifacts in some $\hav$ models.

We note that owing to the remapping to a new reference depth scale,
points at a constant optical depth or column-mass density will end
up probing and spanning a range of geometrical depths, implying that
the averages (and statistical properties) with respect to the new
reference depth scale will be qualitatively and quantitatively different
from plain horizontal averages on constant geometrical depth (see
App. \ref{app:Remarks-on-averages}).

\subsection{Extrapolation at the top\label{sub:Extrapolation-at-the-top}}

The vast majority of \textsc{Stagger}-grid models are sufficiently
extended vertically, in particular at the top, to embrace the full
range of $\nltau$ with $\left[-5.0,+5.0\right]$. The condition $\left\langle \ltaur\right\rangle _{\mathrm{top}}\leq-6.0$,
is usually fulfilled for all but a few models. More specifically,
surfaces of constant optical depth can become quite corrugated at
the top for some giant models and fall outside the physical domain
of the simulations; that is, one can occasionally have $\ltaur^{\mathrm{top}}>-5.0$
for a limited number of columns. These particular columns are therefore
linearly extrapolated to $\ltaur=-5.0$ to allow calculating of average
quantities in the desired range of optical depths. Exponentially varying
values like density, pressure opacities are extrapolated by considering
their logarithmic values. The extrapolation is needed only for a few
giant models ($\logg\leq2.5$), and the concerned columns are usually
only a small fraction ($\lesssim0.3\%$). Therefore, we regard these
extrapolations as negligible in the case of the optical depth scale
$\taur$. 

For the optical depth scale $\tau_{500}$, the situation is slightly
different. The mean optical depth at 500~nm at the top $\left\langle \log\tau_{500}\right\rangle _{\mathrm{top}}$
deviates increasingly towards giant models from $\left\langle \ltaur\right\rangle _{\mathrm{top}}$,
so that $\left\langle \log\tau_{500}\right\rangle _{\mathrm{top}}>-5.0$.
Therefore, the necessary extrapolation at the top is considerable,
in particular for giant models.

We notice that careless column-wise extrapolation at the top can lead
to a largely uncertain and erroneous stratification, which would have
a negative impact on spectral line formation. For instance, a wrong
density stratification at the top can dramatically affect the ionization
balance. To limit these extrapolation errors, we first restrict the
column-wise extrapolation to the region $\nltau_{\mathrm{500}}\geq\nltau_{\mathrm{top}}$
where the value $\nltau_{\mathrm{top}}>-5.0$ is chosen so that no
more than $20\%$ of the columns would require extrapolation up to
that level. We then compute the horizontal averages (step 3) and,
after that, linearly extrapolate the $\hav$ models a second time
to the original $\nltau_{\mathrm{top}}=-5.0$ for each time snapshot.
This particular extrapolation procedure produces more plausible stratifications
since the horizontal $\hav$ averages exhibit a smooth and monotonic
behavior with depth at the top compared to individual columns of the
3D data set. 

Test calculations of data sets from the solar simulation, which were
truncated at the top, revealed the reliability of this \textit{double
extrapolation} approach, since for the temperature stratifications
we find the maximum error around $1\,\%$ at the top ($\nltau_{\mathrm{top}}=-5.0$).
Nonetheless, we favor the use of averages on mean Rosseland optical
depth, i.e. $\havr$ rather than $\havf$, since these averages are
not plagued by such extrapolation uncertainties. For the extrapolated
models on $\tau_{500}$, we kept track of the extent of the applied
extrapolation; in fact, only a few models with the lowest gravities
($\logg=1.5/2.0$) exhibit a noteworthy extrapolation ($\nltau_{\mathrm{top}}\simeq-4.3/4.8$,
respectively). The $\havf$ averages can therefore be reduced to the
extrapolation-free regime at the top afterwards.

\section{Comparison of the averaging methods\label{sec:Comparison-of-averages}}

In the following, we systematically compare the different types of
averaging procedures explained in Sect. \ref{sec:Methods} over a
broad range of stellar parameters relative to Rosseland optical depth,
i.e. $\hav_{\tilde{z}}-\havr$. For the sake of clarity, we illustrate
the properties of average stratifications only for a representative
selection of \textsc{Stagger}-grid models comprising dwarfs and giants
($\logg=4.5$ and $2.0$) at solar and subsolar metallicity ($\feh=0.0$
and $-3.0$). Besides the most important thermodynamic state variables,
temperature and density, we also investigate averages of electron
number density, an important quantity for, say, calculations of ionization
balance and spectral line formation.

Owing to the lack of a unique common global depth scale that is invariant
between different averaging methods, we display their results jointly
on the averaged Rosseland optical depth scale, $\left\langle \taur\right\rangle $,
in order to enable a direct comparison.

\subsection{Temperature\label{sub:Temperature}}

\begin{figure*}
\includegraphics[width=88mm]{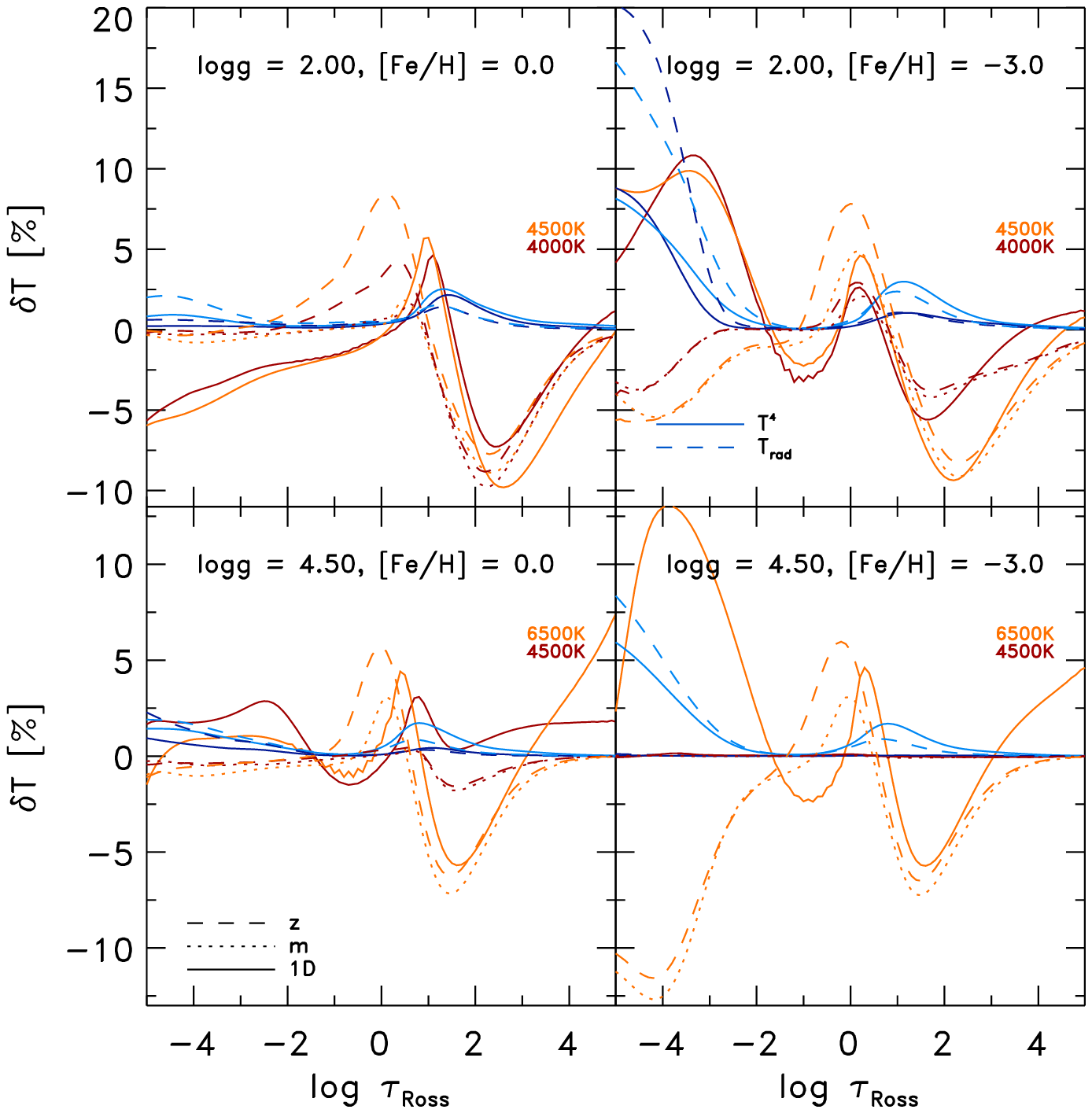}\includegraphics[width=88mm]{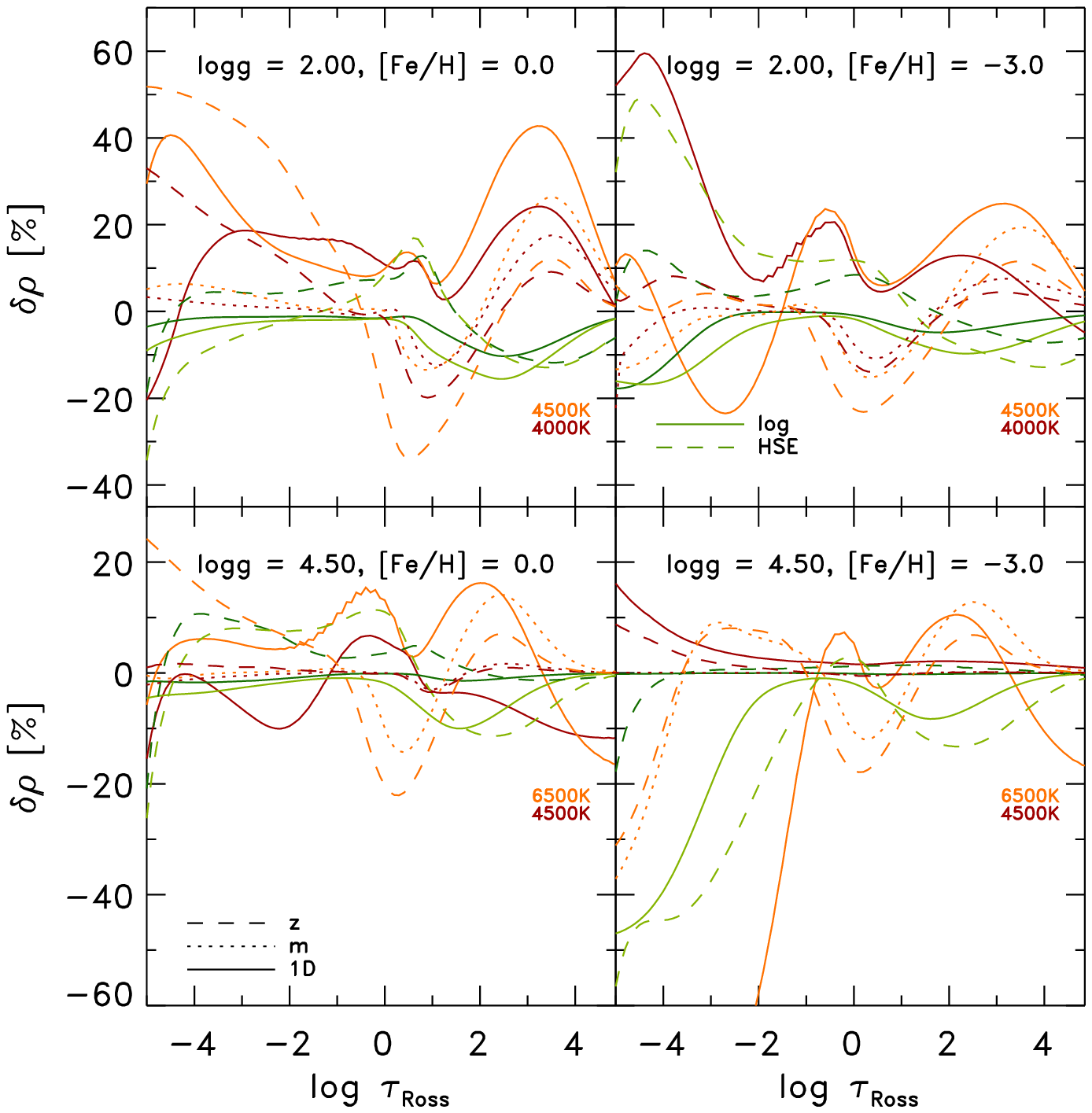}

\caption{Relative differences in the temperature (left) and density (right
panel) stratification vs. the (averaged) Rosseland optical depth for
various stellar parameters. The differences are relative to the Rosseland
optical depth, i.e. $\hav_{\tilde{z}}-\havr$.\emph{ Orange/brown
dashed lines}: averages on layers of constant geometrical height $\havz$;
\emph{orange/brown dotted lines}: averages on layers of constant column
mass density $\havcm$; \emph{orange/brown solid lines}: 1D MLT models.
\emph{Blue solid lines}: flux-weighted $T^{4}$-stratifications; \emph{blue
dashed lines}: brightness temperatures $T_{\mathrm{rad}}$ averaged
on surfaces of constant Rosseland optical depth (left panel). \emph{Green
solid lines}: logarithmic density averages $\havr^{\mathrm{log}}$;
\emph{green dashed lines}: hydrostatic averages $\havr^{\mathrm{HSE}}$
(right panel). We compare always cooler and hotter effective temperatures,
which are distinguished by dark and bright colors respectively. We
note that the cool metal-poor dwarfs exhibit very small differences,
and are therefore indistinguishable. Note the differences in the $y$-axes.}

\label{fig:temp}\label{fig:density} 
\end{figure*}
We find that the temperature stratifications of the two optical reference
depth scales, $\havr$ and $\havf$, are similar, therefore we refrain
from showing these. Only at the top of the metal-poor stars do the
$\havf$-averages appear cooler ($\sim5\,\%$, i.e by $\gtrsim250\,\mathrm{K}$
at $\teff=6000\,\mathrm{K}$). On the other hand, the geometrical
$\havz$ and column mass density $\havcm$ averages deviate distinctively
from the $\havr$-stratification (see Fig. \ref{fig:temp}). In the
regime $1.0<\ltaur<3.0$, both $\havz$ and $\havcm$ are cooler by
$\sim5-10\,\%$. At the surface ($\tau_{\mathrm{Ross}}=0$), the geometrical
averages deviate considerably, while the $\havcm$-averages are closer
to the optical depth scale (see Fig. \ref{fig:temp}). In the deeper
layers below the superadiabatic regime (SAR), the various averaging
methods are practically indistinguishable. In the upper atmosphere
the differences are smaller at higher $\feh$ due to relatively low
horizontal contrast, but, these increase significantly for lower metallicity.
The averages $\havz$ and $\havcm$ are marginally cooler than $\hav_{\mathrm{Ross}}$
by $\sim1-2\,\%$ at solar metallicity. In the metal-poor case $\feh=-3.0$,
the temperature stratifications are distinctively cooler, which will
certainly influence the line formation calculations with $\hav$ stratifications.
Furthermore, the differences increase with higher $\teff$ and lower
$\logg$.

As mentioned earlier, in the atmospheres of late-type stars, minor
temperature fluctuations are amplified disproportionally into large
variations in the line and continuum opacity $\kappa_{\lambda}$ owing
to the strong $T$-sensitivity of the $\mathrm{H}^{-}$-opacity ($\kappa_{\lambda}{\propto}T^{10}$,
see \citealt{Stein:1998p3801}). Therefore, surfaces of constant optical
depth appear strongly corrugated in terms of the range of geometrical
heights that they span. The transformation to layers of constant optical
depth will naturally even out these corrugated surfaces and, at the
same time, smooth the temperature fluctuations, since the latter are
the source of the former (see App. \ref{app:Reversed-granulation}).
Therefore, these are noticeably smaller on layers of constant optical
depth compared to layers of constant geometrical depth, which is portrayed
in the temperature contrast and histogram (see also Figs. \ref{fig:tt_cont}
and \ref{fig:tt_hist}). The SAR exhibits large-amplitude fluctuations
as a result of the release of thermal and ionization energy at the
photospheric transition, which are the reason for the observed enhanced
differences between the averaging methods (see Sect. \ref{sub:Contrast}).

\citet{Steffen:2002p18843} found a beneficial mean $\left\langle T\right\rangle $-representation
for the Sun in the flux-weighted temperature averages, $T^{4}$, taken
on constant Rosseland optical depth from their 2D simulations. The
idea behind this approach is that the $T^{4}$-averages render radiation-oriented
$T$-stratifications, therefore resulting in 1D line profiles that
are closer to the multidimensional ones \citep[see also][]{Steffen:1995p14024}.
To allow for a similar comparison for our models, we computed such
average $T^{4}$-stratifications. In Fig. \ref{fig:temp}, the $T_{\mathrm{Ross}}^{4}$-stratifications
generally appear hotter at the top and in the SAR compared to the
simple $T$-stratification. Averages taken at the fourth power will
weight higher values more, which leads to hotter average temperatures.
This could lead to pronounced differences for molecular lines that
form high up in the atmosphere. At solar metallicity, the $T^{4}$-stratifications
at the top are fairly similar to the plain $T$-averages ($\sim1-2\,\%$)
in agreement with the findings of \citet{Steffen:2002p18843}. This
is different at lower metallicity ($\feh=-3.0$), namely the $T^{4}$-averages
are clearly higher by $\sim5-10\,\%$. At higher $\teff$ and lower
$\logg$, the temperature differences are greater, in particular for
the metal-poor giants, owing to the enhanced temperature fluctuations
(see Sect. \ref{sub:Contrast}).

Under the assumption of local thermodynamic equilibrium (LTE) and
neglecting the effects of scattering, the source function is given
by the Planck function, $S_{\lambda}=B_{\lambda}\left(T\right)$.
Within this approximation, we can thus consider the brightness temperature
average $T_{\mathrm{rad}}$ defined earlier in Sect. \ref{sub:aAveraging}
as a good representation of the mean temperature stratification from
the point of view of the radiative emission properties: brighter parts
in each depth layer are given more weight with this averaging method.
The differences between the average $T_{\mathrm{rad}}$ at $500\,\mathrm{nm}$
and average $T$-stratifications are displayed in Fig. \ref{fig:temp}.
Their variations with stellar parameters are very similar to those
of $T^{4}$-averages, however, slightly more pronounced, in particular
the metal-poor giants exhibit hotter stratifications by up to $\sim20\,\%$
at the top.

\subsection{Density\label{sub:Density}}

In Fig. \ref{fig:density}, we also illustrate the results of averaging
in the case of the density stratifications. In the deeper interior,
the different $\hav$ models converge toward the same density stratification.
In the SAR, below the optical surface at $\ltaur\gtrsim0.0$, the
geometrical averages $\havz$ are smaller than the $\havr$ averages
by up to $\sim30\,\%$, while at the top these are much denser by
up to $\sim40\,\%$. The differences increase towards higher $\teff$
and lower $\logg$. We find a different behavior in the metal-poor
dwarfs, which turn lower towards the top after the initial increase
($\sim10\,\%$). The density stratifications averaged on column mass
density $\havcm$ are larger in the SAR and in the upper layers closer
to $\havr$. However, we find that at lower metallicity $\left\langle \rho\right\rangle _{m}$
they are smaller by up to $\sim30\,\%$. We note that thermal pressure
qualitatively shows the same characteristics as the density.

The shape of the density distribution is symmetric and narrow on layers
of constant column mass density, thanks to the exponential stratification
of the atmosphere and to the additional damping of density fluctuations
on the column mass scale (see Fig. \ref{fig:rho_hist}). As a result,
the $\havcm$ averages feature the narrowest contrast and density
ranges, which, on the contrary, are usually greatest for geometrical
averages $\havz$; for the $\havr$ averages, these are noticeably
reduced due to the mapping onto the optical reference depth scale
(Fig. \ref{fig:rho_cont}). Overall, the density fluctuations at the
top of the $\havr$ stratifications are similarly as small as those
by $\havcm$ and $\sim20\,\%$; however, for metal-poor dwarfs they
reach up to $\sim80\,\%$ (see Fig. \ref{fig:tt_cont}). As shown
in Sect. \ref{sub:Histograms}, we find that the corrugation of the
layers of constant optical depth in the upper part of 3D model stellar
atmospheres at lower metallicity increases considerably towards higher
$\teff$ because of an enhanced $T$-contrast by the so-called reversed
granulation \citep[see][]{Rutten:2004p16166}. This in turn broadens
the density distribution during the remapping to the optical depth
scale, shifting the mean density value and leading to the observed
deviations between $\left\langle \rho\right\rangle _{\mathrm{Ross}}$
and $\left\langle \rho\right\rangle _{m}$ at lower metallicity (see
App. \ref{app:Reversed-granulation}), which will affect the $\hav$
line formation calculations.

The highly stratified structure of stellar atmospheres features an
exponential decrease with height. Linear density averages will therefore
tend to give more weight to higher density values, leading to a systematic
overestimation of the mean densities. For this reason we consider
the logarithmic averages $\left\langle \rho\right\rangle _{\mathrm{log}}$,
which we compare to the linear ones in Fig. \ref{fig:density}. As
expected, we find the logarithmic $\rho$-averages are smaller than
the linear ones, with the difference between the two increasing with
higher $\teff$ and lower $\logg$ by up to $\sim30\,\%$. The mean
densities in the upper layers are lower by $\sim10\,\%$ and $\sim40\,\%$
at solar and low metallicity, respectively. For quantities that vary
more moderately (e.g., temperature) the differences between logarithmic
and linear averaging are rather small.

The transformation to constant optical depth and the subsequent averaging
will change the physical self-consistency as shown in App. \ref{app:hse_stratification}.
To rectify this, we followed the recommendation of \citet{Uitenbroek:2011p10448}
and also computed $\rho$-stratifications, which are enforced to be
in hydrostatic equilibrium, $\left\langle \rho\right\rangle _{\mathrm{HSE}}$
(Fig. \ref{fig:density}). These deviate significantly from the plain
$\left\langle \rho\right\rangle $-stratifications, in particular
at the top. Incidentally, we note however that their dynamic nature
and the effects of convective flows and turbulent pressure mean that
the 3D models themselves are not strictly speaking in hydrostatic
equilibrium at any one time.

In Fig. \ref{fig:temp} (both panels), we also compare the 1D MLT
models with the $\havr$ stratifications. The 1D models in general
show qualitatively similar behavior as the geometrical averages. The
metal-poor 1D models are distinctively hotter, since these enforce
radiative equilibrium in the upper layers.

\subsection{Electron number density\label{sub:Electron-number-density}}

\begin{figure*}
\includegraphics[width=88mm]{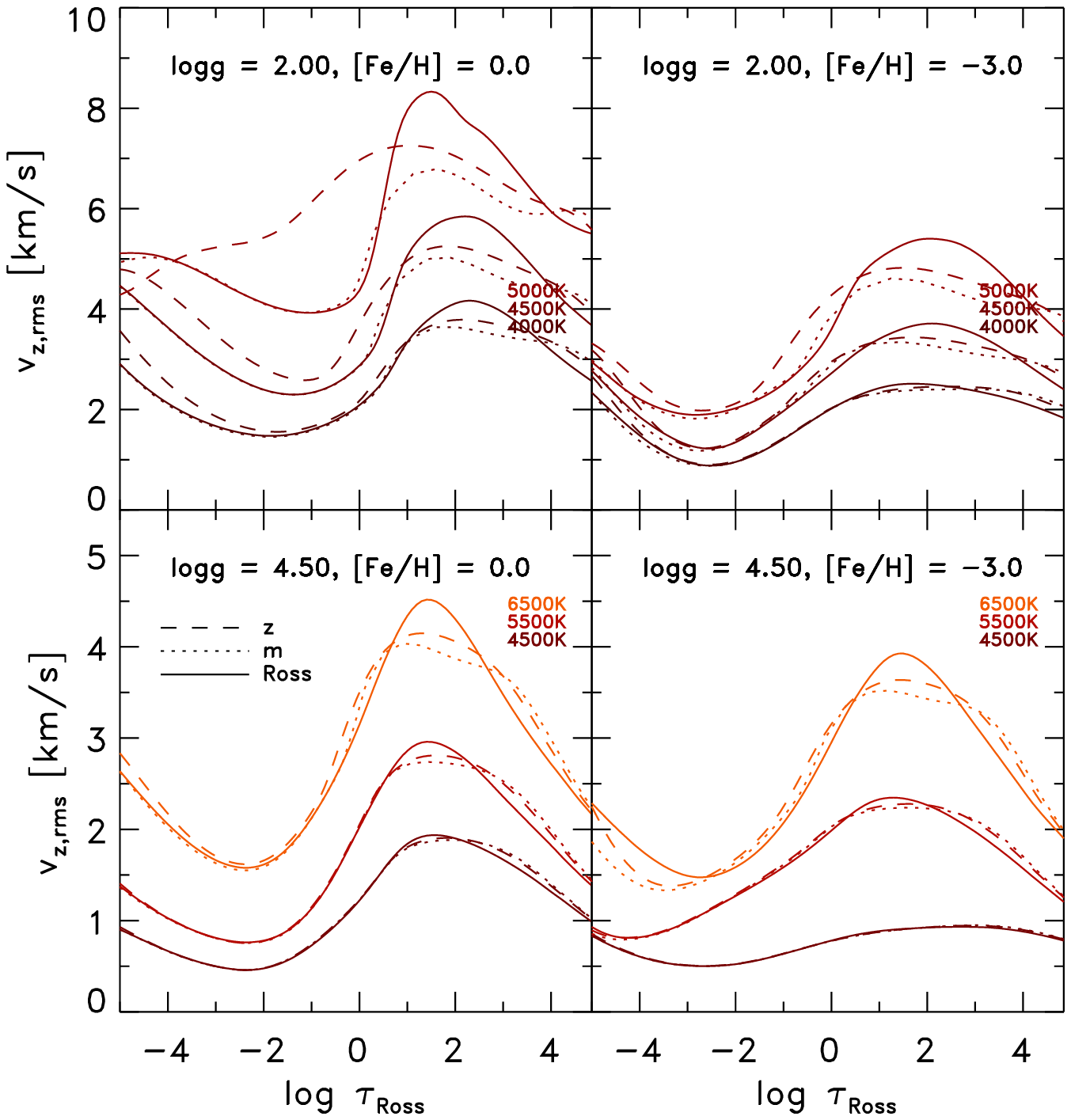}\includegraphics[width=88mm]{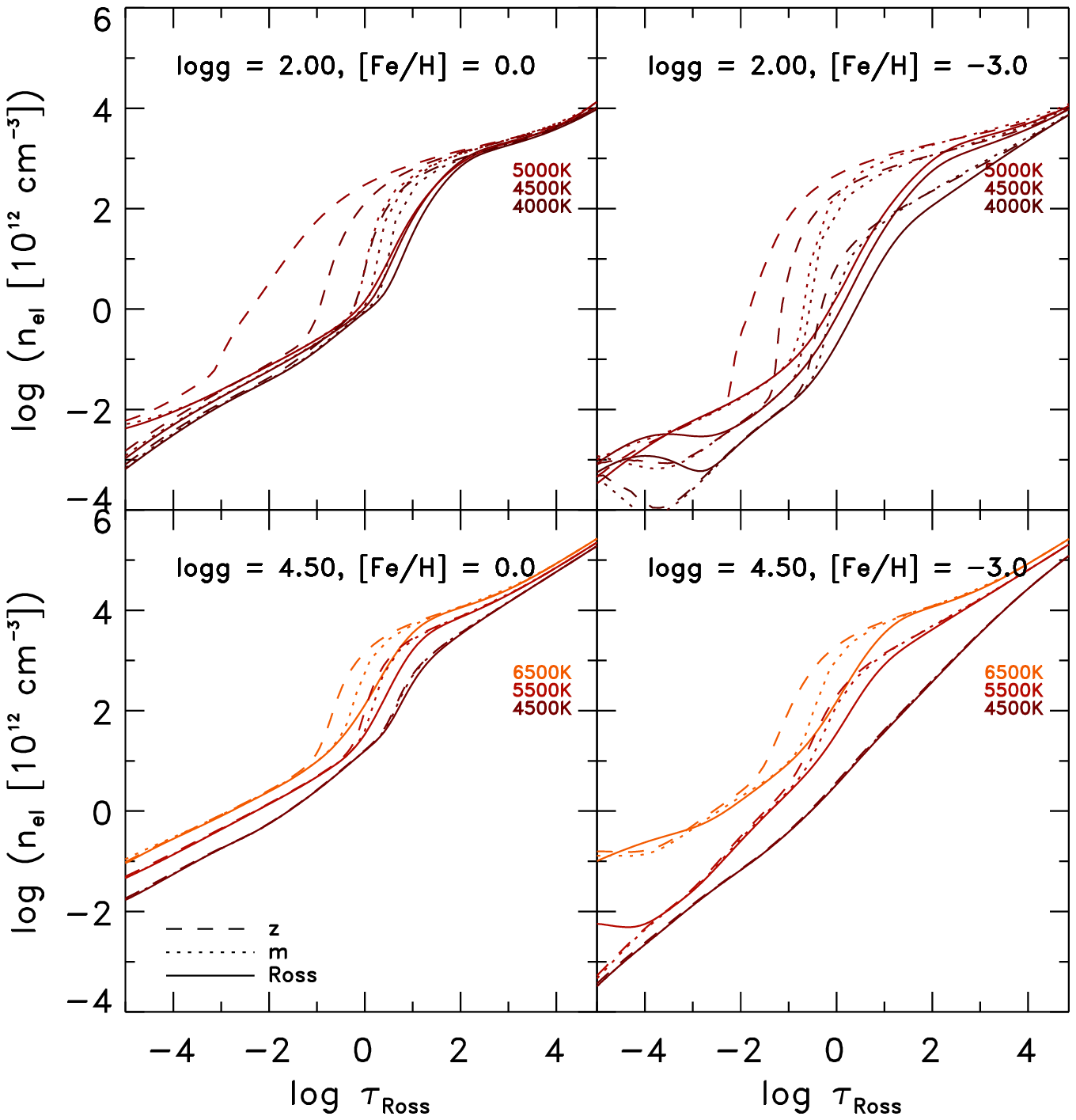}
\caption{Root mean square (rms) of the vertical velocity $\vzrms$ (left) and
mean electron number density $n_{\mathrm{el}}$ vs. optical depth
(right panel). \emph{Dashed lines}: $\havz$ averages; \emph{dotted
lines}: $\havcm$; \emph{solid lines}: $\havr$.}

\label{fig:velocity}\label{fig:electron-density} 
\end{figure*}
We find large differences among the various averages of the electron
number density, $n_{\mathrm{el}}$, which we show in Fig. \ref{fig:electron-density}
(right panel). In the SAR the geometrical averages $\left\langle n_{\mathrm{el}}\right\rangle _{z}$
are distinctively larger than the averages on surfaces of constant
Rosseland optical depth $\left\langle n_{\mathrm{el}}\right\rangle _{\mathrm{Ross}}$,
while the column mass density averages $\left\langle n_{\mathrm{el}}\right\rangle _{m}$
are found in between the two. The deviations increase for higher $\teff$
and lower $\logg$ considerably, while at lower $\teff$ the differences
are significantly smaller. We show in App. \ref{app:Reversed-granulation}
that the interpolation to a new reference depth scale changes the
statistical properties by redistributing properties from different
heights, so the resulting mean horizontal average will look different
depending on the reference depth scale. This effect seems to be most
pronounced in the case of electron density.

To determine the ionization fraction in spectral line calculations,
the electron number density is either already provided by the model
atmosphere or looked up from an EOS using the independent thermodynamic
variables (typically $(T,p)$ or $(T,\rho)$). The latter has to be
done carefully in the case of the $\hav$ models, since, besides potential
differences in the EOS compared to the one used for calculating the
model atmosphere, electron densities derived from the EOS based on
averaged independent variables, $n_{\mathrm{el}}^{\mathrm{EOS}}=n_{\mathrm{el}}\left(\left\langle T\right\rangle ,\left\langle p\right\rangle \right)$,
can deviate significantly from the more physically consistent averaged
$\left\langle n_{\mathrm{el}}\right\rangle $ (see App. \ref{app:Deviations-from-EOS}).

\subsection{Vertical velocity\label{sub:Vertical-velocity}}

It is worthwhile to compare how the vertical velocity, $\vzrms$,
changes with the respective averaging methods. For comparison, we
show in Fig. \ref{fig:velocity} (left panel) the rms of the vertical
velocity. In the upper layers, we find the $\vzrms$ on geometrical
averages to be higher compared to other averages, while it is lower
in the deeper layers. On optical depth the peak in $\vzrms$ below
the surface is somewhat symmetric and slightly higher, while for averages
on geometrical height and column mass density their peaks are flatter
and more skewed towards higher layers, and the peak location is realized
in slightly upper layers. For lower $\teff$ and higher $\logg$,
the differences diminish more and more, so that for the coolest models,
the difference are small. The differences in the velocity arise as
well due to the redistribution of velocity during the mapping to the
new reference depth scale (see App. \ref{app:Reversed-granulation}).

\section{Statistical properties\label{sec:Statistical-properties}}

To explore the origins of the differences among the various average
$\hav$ structures and the resulting ramifications for line formation
calculations, we discuss here the statistical properties of the temperature,
density, and velocity stratifications. Since the statistical properties
of $\havf$ and $\havr$ are fairly similar, we focus only on the
latter.

\subsection{Contrast\label{sub:Contrast}}

\begin{figure}
\includegraphics[width=88mm]{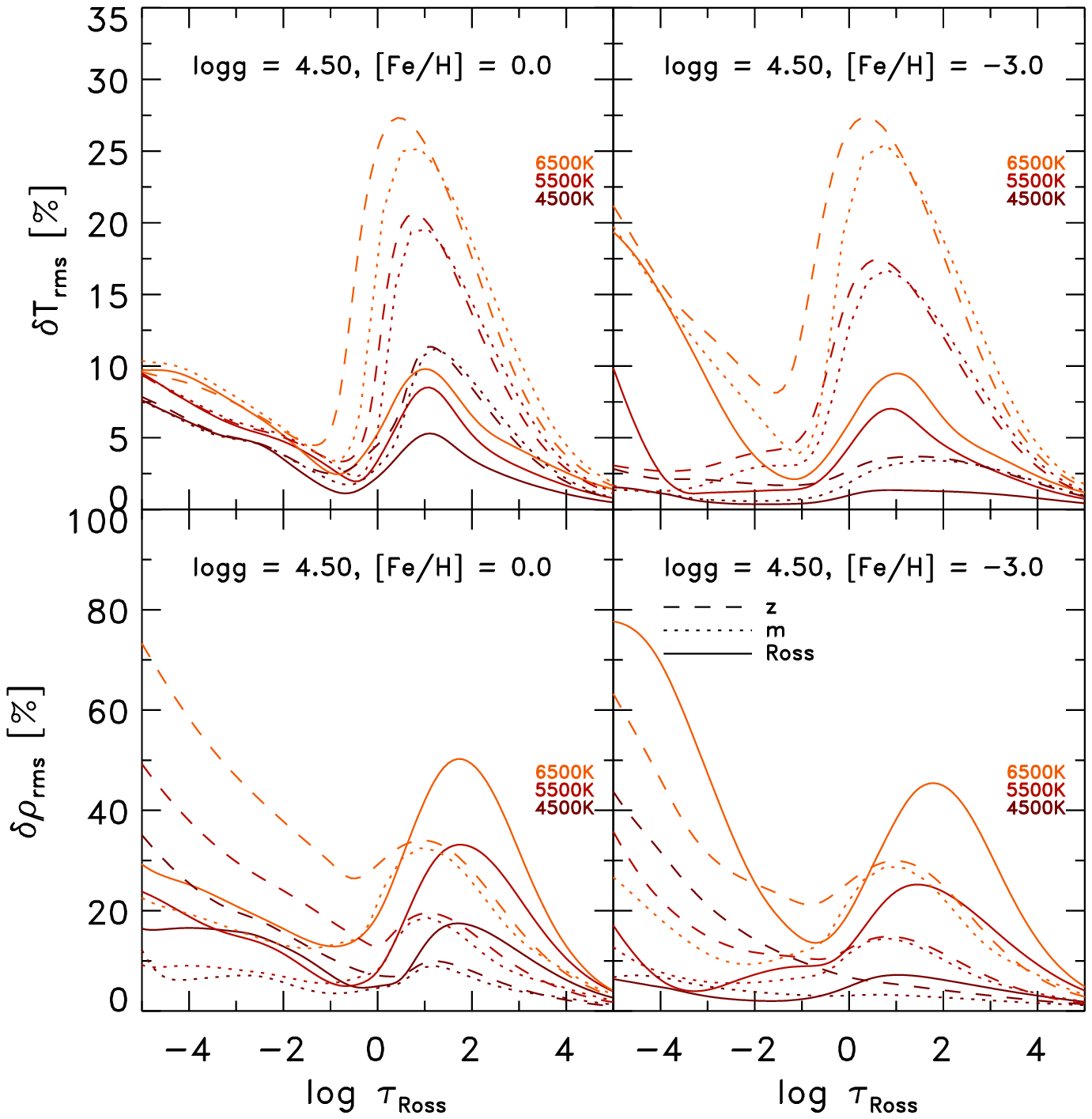}

\caption{Temperature (top) and density (bottom) contrasts vs. averaged Rosseland
optical depth\emph{. Dashed lines}: $\havz$ averages; \emph{dotted
lines}: $\havcm$; \emph{solid lines}: $\havr$.}

\label{fig:tt_cont}\label{fig:rho_cont}\label{fig:tt_ext}\label{fig:rho_ext} 
\end{figure}
The 3D RHD models usually exhibit a broad range of values at a given
height thanks to the fluctuations arising from the convective motions.
The amplitude of these fluctuations can be quantified using the root-mean-square
of the relative deviation from the mean, 
\begin{equation}
\delta X_{\mathrm{rms}}=\sqrt{\Sigma_{i=1}^{N}\left(X_{i}-\bar{X}\right)^{2}/\left(N\bar{X}^{2}\right)},\label{eq:contrast}
\end{equation}
which we refer to as the \emph{contrast} ($\bar{X}$ is the mean value
of $X$). It is equal to the normalized standard deviation; i.e.,
$\delta X_{\mathrm{rms}}=\sigma_{X}/\bar{X}$.

The translation to another reference depth scale changes the statistical
properties as variables are remapped, which in turn is reflected in
changes in contrast. Among the various averaging methods, geometric
averages $\havz$ typically feature the highest contrast. We also
find that the level of fluctuations generally increases with increasing
$\teff$ and decreasing $\logg$. The highest contrast typically prevails
in simulations with the highest $\teff$ and located in the vicinity
of the maximum superadiabatic gradient, $\nsadp$, and maximum rms-velocity,
$\vzrmsp$. These arise from the photospheric transition from convective
to radiative energy transport, and the resulting overturning of the
entropy-depleted plasma. At the top of the convection zone, the fluctuations
reach a minimum, and they decrease towards the bottom of the model
atmosphere.

In top and bottom panels of Fig. \ref{fig:tt_cont}, we show the temperature
and density contrasts, $\delta T_{\mathrm{rms}}$ and $\delta\rho_{\mathrm{rms}}$,
respectively. In the case of the optical depth $\havr$, the temperature
contrast is significantly reduced compared to the other reference
depth scales ($\delta T_{\mathrm{rms}}^{\mathrm{peak}}$ reduced by
a factor of $\sim3$), while the density contrast is slightly enhanced
($\delta\rho_{\mathrm{rms}}^{\mathrm{peak}}\sim20-60\,\%$ compared
to $10-50\,\%$). For averages on column mass density $\havcm$, $\delta\rho_{\mathrm{rms}}$
is lower, in particular in the upper layers, and $\delta T_{\mathrm{rms}}$
is slightly smaller compared to the $\havz$ case. Fluctuations of
variables that correlate with the new reference depth scale will be
reduced during the transformation. As the translation to layers of
constant optical depth partly evens out the corrugated $\tau$-isosurface,
fluctuations of the opacity $\kappa_{\lambda}$ will be reduced, since
the dominant $\mathrm{H}^{-}$opacity is very sensitive to temperature.
Therefore, the temperature fluctuations are also smoothed out. Layers
of constant column mass density will similarly suppress density variations
(see App. \ref{app:Reversed-granulation}). At the top, $\delta\rho_{\mathrm{rms}}$
is almost similar between $\havcm$ and $\havr$ in the case of the
solar metallicity $(\delta\rho_{\mathrm{rms}}^{\mathrm{top}}\sim40\,\%$);
however, at lower metallicity, $\feh=-3.0$, we find considerable
disparity with $\delta\rho_{\mathrm{rms}}^{\mathrm{top}}\sim80\,\%$.

The thermal stratification in the upper atmosphere is determined by
adiabatic cooling thanks to mechanical expansion and radiative heating
because of spectral line re-absorption \citep{Asplund:1999p11771,Collet:2007p5617}.
In metal-poor stars, radiative reheating in upper layers is significantly
reduced owing to the weakness of spectral line features, while the
mechanical expansion cooling term is virtually unaffected. The reversed
granulation takes place at increasingly lower geometrical height with
higher $\teff$ and lower $\logg$, causing the distribution of the
thermodynamic variables to become increasingly broader and more skewed
(see Sect. \ref{sub:Histograms}). This is the reason for the enhancement
in $\delta T_{\mathrm{rms}}$ and $\delta\rho_{\mathrm{rms}}$ towards
the top boundary in metal-poor simulations in Fig. \ref{fig:tt_cont}.
Replicating the results of full 3D line formation calculations in
low-metallicity stars with $\hav$ models is therefore challenging,
since the averages have to correctly account for such temperature
and density fluctuations. Interestingly, the temperature contrast
saturates at $6500\,\mathrm{K}$, similar to the saturation of the
intensity contrast shown in our previous work (see Fig. 10 in Paper
I).

The strength of spectral lines is sensitive to temperature, and the
remapping to constant optical depth decreases $\delta T_{\mathrm{rms}}$,
making $\left\langle T\right\rangle $ closer to $\left\langle T\right\rangle _{\mathrm{rad}}$.
However, the transformation to layers of constant optical depth exhibits
the side effect of redistributing the other variables, too, in particular
the gas density; $\rhoc$ is thus much higher than averages on column
mass density, due to the additional influence of opacity on the depth
scale (see Sect. \ref{sub:aAveraging}). This in turn will likely
affect the line formation calculations with the different $\hav$
models. 

The strong contrast in the upper part of the convection zone ($\ltaur\ge0$)
is induced by the large amplitude fluctuations owing to the radiative
energy losses at the photosphere and the asymmetry of the up- and
downflows, which we discuss further in Sect. \ref{sub:Up-and-downflows}.
An interesting aspect is that the contrast in thermodynamic variables
is very similar to the rms of the vertical velocity (Fig. \ref{fig:velocity}),
which is indicative of the correlation between the mass flux and the
fluctuations in the thermodynamic variables. Namely, vertical velocity
is generated by density contrast $\delta\rho$ via to the buoyancy
force, $f_{B}=-g\delta\rho$, which results from an imbalance of pressure
and gravity terms in the hydrodynamical equation for conservation
of momentum (see Paper I) in the highly stratified atmosphere. Lighter
fluid elements ($\delta\rho<0$) experience positive buoyancy and
thus upward acceleration, while denser elements ($\delta\rho>0$)
experience negative buoyancy and are pulled downward. Buoyancy forces
will vanish eventually, when the density of the up- or downflowing
element levels with the surrounding gas.

The entropy contrast $\delta s_{\mathrm{rms}}$ (not shown here),
qualitatively depicts a very similar dependence on stellar parameter
and reference depth scale as $\delta T_{\mathrm{rms}}$. Both are
very similar in optical depth, while for the averages $\havz$ and
$\havcm$ the overall amplitude is a factor $\sim2$ smaller. In Paper
I, we showed that the convective energy flux depends on the entropy
jump, density, and vertical velocity. Interestingly, here we also
find additional \emph{scaling relations} concerning the peak contrast
in entropy, $\delta s_{\mathrm{rms}}^{\mathrm{peak}}$, and density,
$\delta\rho_{\mathrm{rms}}^{\mathrm{peak}}$, with the vertical peak
velocity $v_{z,\mathrm{rms}}^{\mathrm{peak}}$. This can be interpreted
as convective driving, where the radiative losses generate large fluctuations
in the entropy, temperature, and density.

For the different averaging methods, the variations in the minimum-maximum
range for the temperature and density are qualitatively very similar
to the contrast (even though with larger amplitudes $\sim5-8$), therefore,
we refrain from discussing these explicitly.

\subsection{Upflows and downflows\label{sub:Up-and-downflows}}

\begin{figure}
\includegraphics[width=88mm]{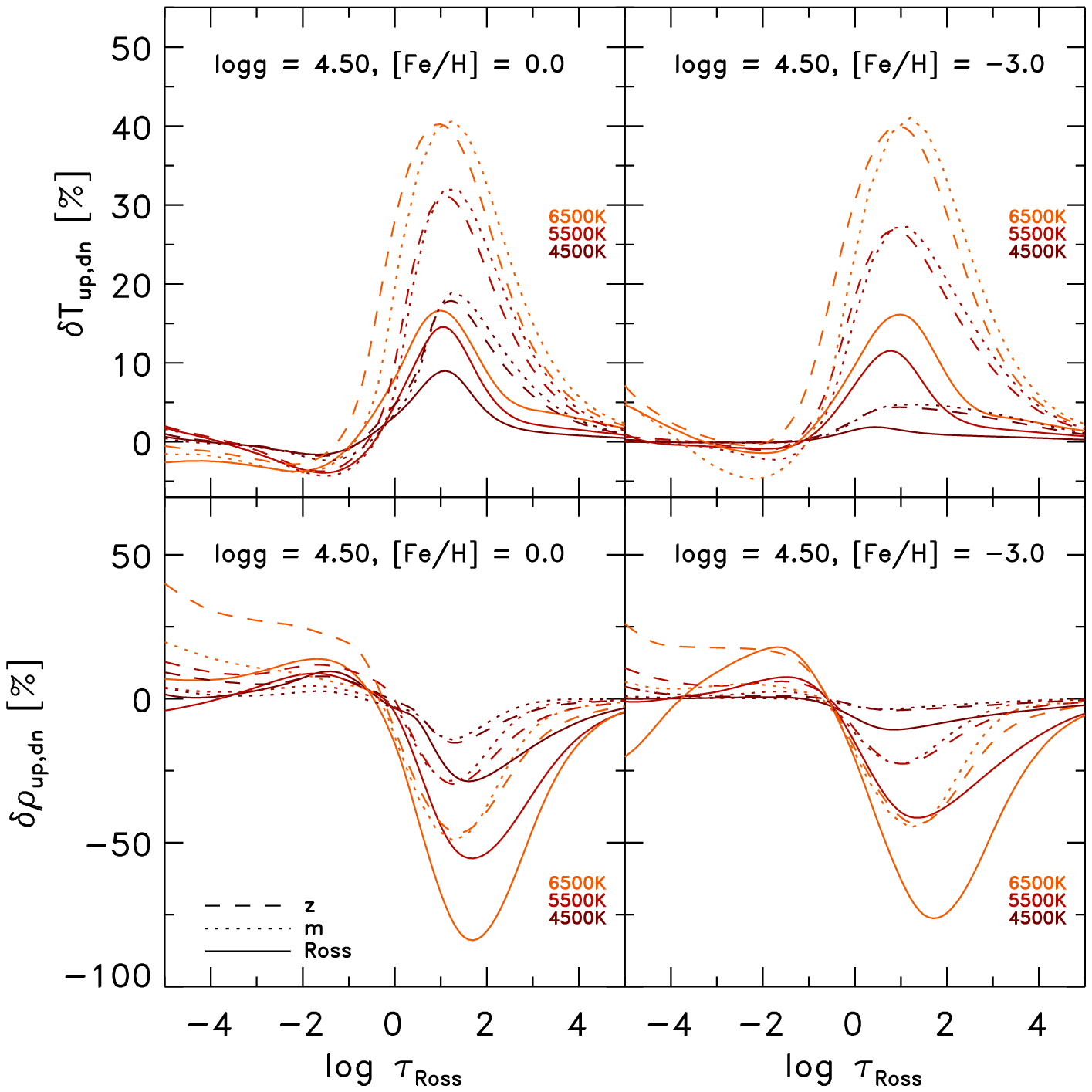}

\caption{Similar as Fig.~\ref{fig:tt_cont} but showing the relative difference
between averages in up and downflows, $\delta T_{\mathrm{up,dn}}$
and $\delta\rho_{\mathrm{up,dn}}$.}

\label{fig:tt_updn}\label{fig:rho_updn} 
\end{figure}
The properties of the convective motions in stellar atmospheres are
highly asymmetric in up- and downflows. The upflows overshoot into
the photosphere leading to non-thermal Doppler shifts imprinted on
spectral line features. We first compute the mean values of various
variables separately for up- and downflows based on the sign of the
velocity at a given height. We then determine the relative difference
between up- and downflows with $\delta X_{\mathrm{up,dn}}=(X_{\mathrm{up}}-X_{\mathrm{dn}})/\bar{X}$
(Fig. \ref{fig:tt_updn}). As expected, the buoyant upflows are hotter
and lighter compared to the subsiding downflows. Furthermore, the
asymmetries are especially pronounced in the convection zone below
the optical surface. Above the photosphere, the convective motions
decay quickly, and the asymmetries in $\delta T_{\mathrm{up,dn}}$
and $\delta\rho_{\mathrm{up,dn}}$ are distinctively smaller. The
remaining asymmetries at the top stem from reverse granulation.

The convective flows in granules, slow and almost laminar, radiate
away their energy and overturn into the intergranular lanes characterized
by cool, dense, narrow turbulent downdrafts. The subsequent large-amplitude
fluctuations in the thermodynamical properties are caused by the turbulent
mixing of the downflows with the upflows, typically located in the
intergranular lane below the optical surface in the SAR. These regions
are arranged in tubelike structures around the granules, and can be
identified with their excessive vorticity. It is remarkable that,
across all stellar parameters, the filling factor of the up- and downflow
in the convection zone remains almost constant, with $f_{\mathrm{up}}\sim2/3$
and $f_{\mathrm{dn}}\sim1/3$, respectively (see Paper I).

The variable $\delta T_{\mathrm{up,dn}}$ is reduced, and $\delta\rho_{\mathrm{up,dn}}$
is enhanced on the optical reference depth scale $\havr$ compared
to the other averages. The column mass density shows a smaller asymmetry
in density. This behavior, similar to what we discussed earlier for
the temperature and density contrasts, is not entirely surprising,
since the fluctuations are caused by the presence of the up- and downflows
(see also \ref{app:Reversed-granulation}).

\subsection{Histograms\label{sub:Histograms}}

\begin{figure}
\includegraphics[width=88mm]{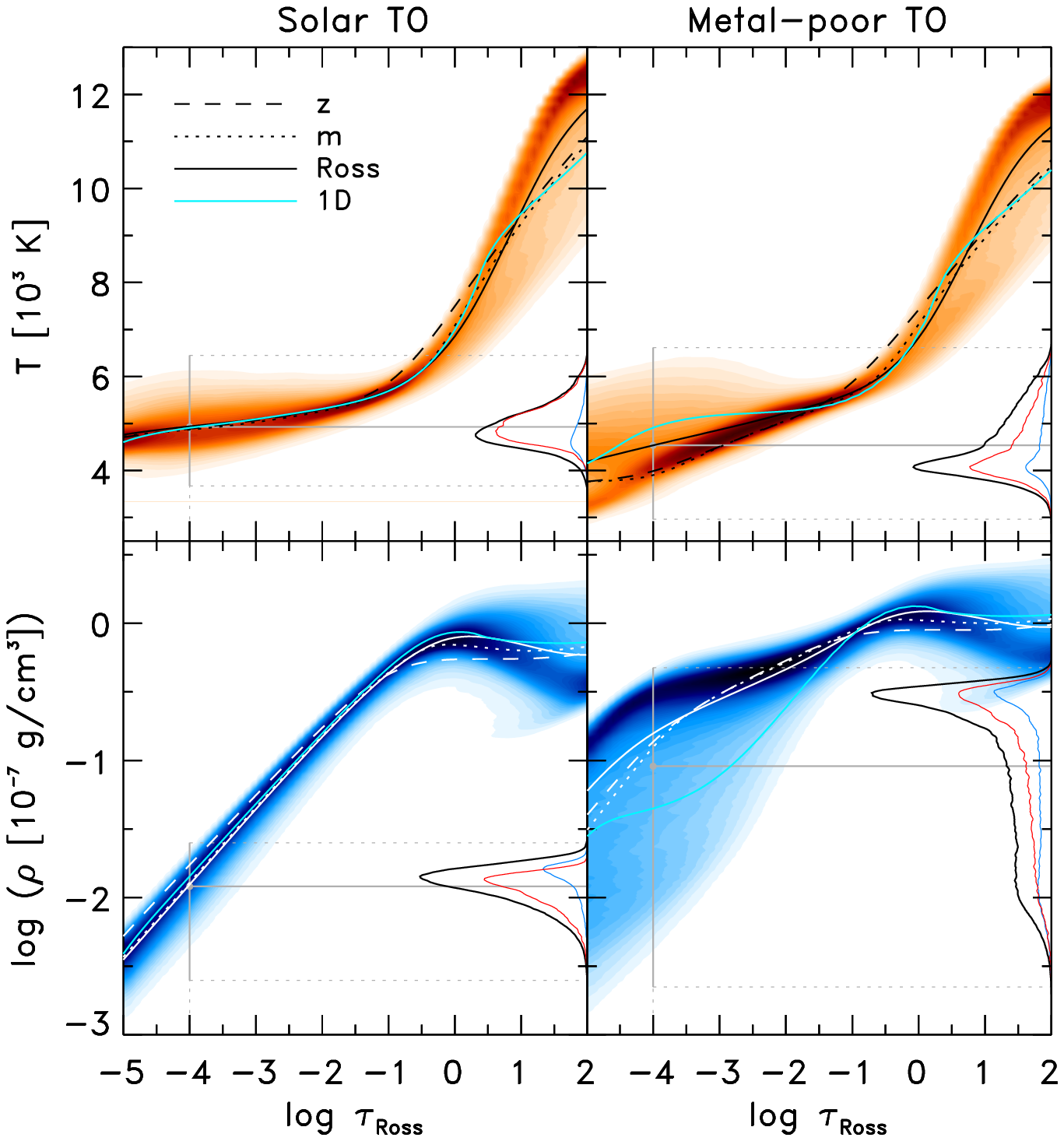}

\caption{Histogram of the temperature (top) and density (bottom) vs. optical
depth for the TO simulation ($\teff=6500\,\mathrm{K}/\logg=4.0$)
with solar and sub-solar metallicity ($\feh=-3.0$). Additionally,
the histogram of a single layer ($\ltaur=-4.0$) is indicated for
the whole layer (black) and separated in up- and downflows (blue and
red, respectively). \emph{Dashed lines}: $\havz$ averages; \emph{dotted
lines}: $\havcm$; \emph{solid lines}: $\havr$; \emph{blue solid
lines}: 1D MLT models.}

\label{fig:hist_to_ov} 
\end{figure}
\begin{figure*}
\includegraphics[width=88mm]{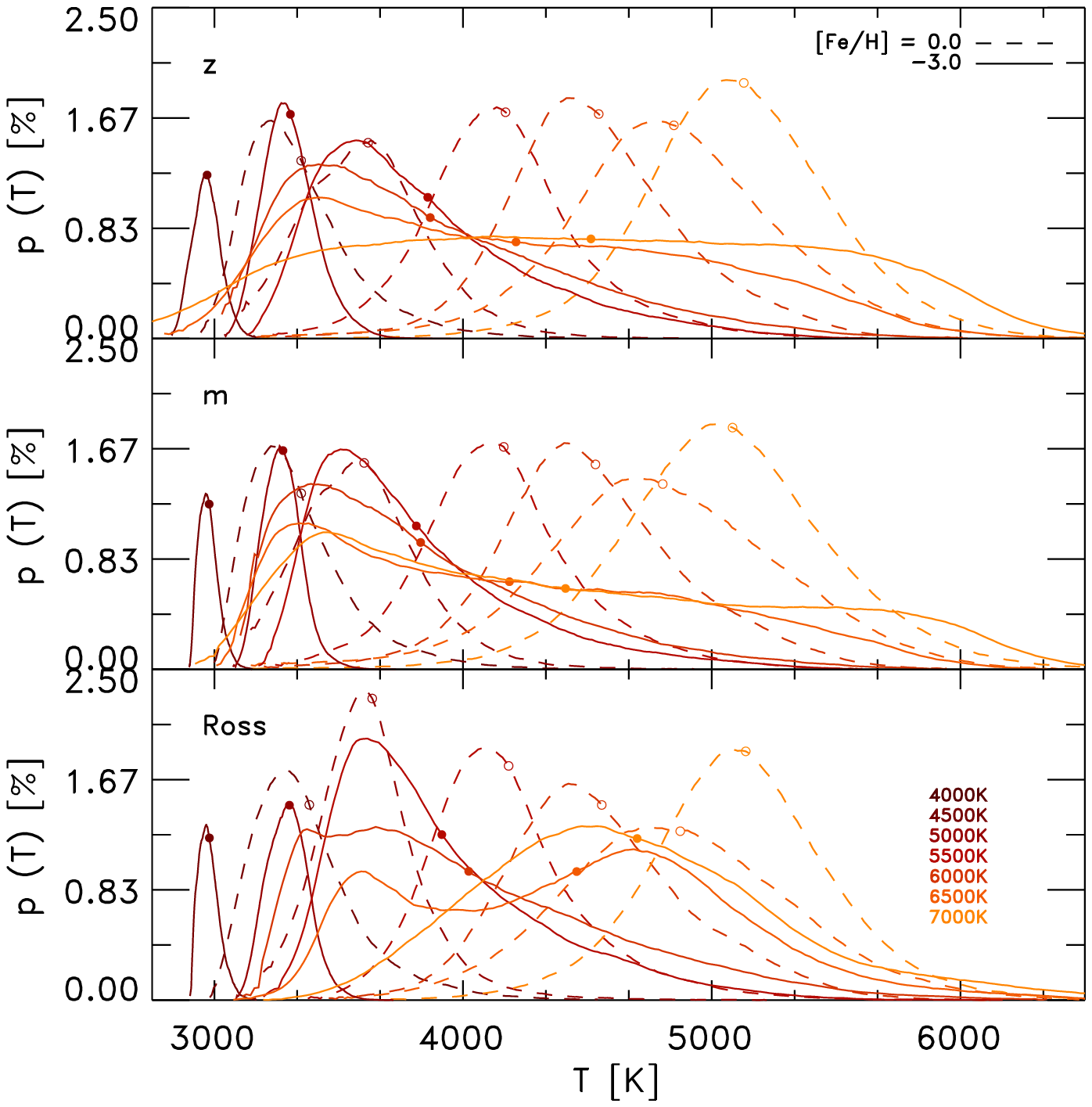}\includegraphics[width=88mm]{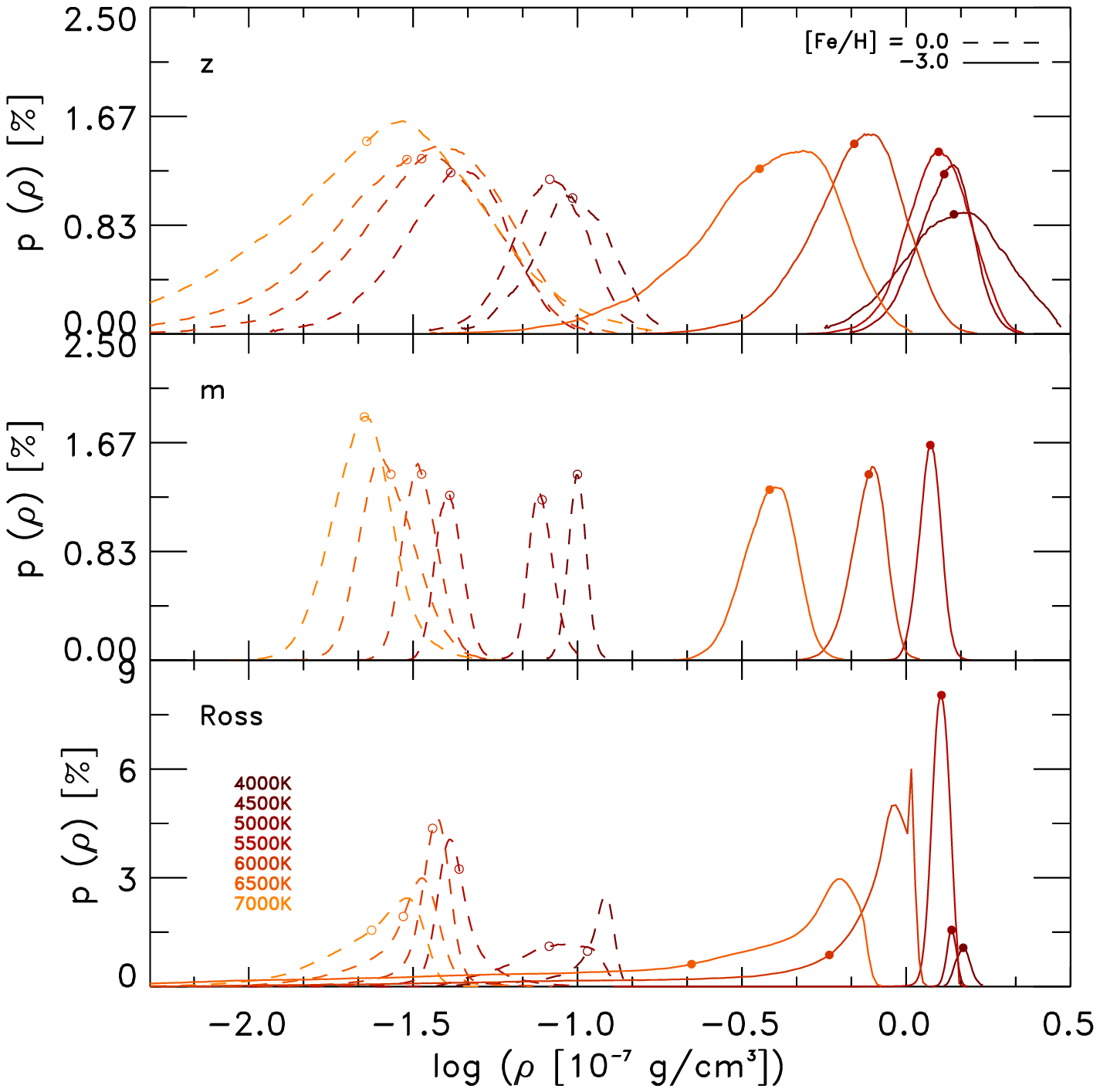}

\caption{Histograms of the temperature (left) and density (right panel) distributions
taken at $\left\langle \ltaur\right\rangle =-4.0$. We show the histograms
averaged on constant geometrical height (top), column mass density
(middle), and Rosseland optical depth (bottom). The surface gravity
of displayed models is $\logg=4.5$ and the metallicity is solar (dashed
lines) and subsolar with $\feh=-3.0$ (solid lines). The mean values
are indicated by filled and open circles for $\feh=-3.0$ and 0.0,
respectively).}

\label{fig:tt_hist}\label{fig:rho_hist}\label{fig:temperature-distribution}\label{fig:density-distribution} 
\end{figure*}
In Fig. \ref{fig:hist_to_ov}, we illustrate temporally averaged histograms
of the temperature, $p\left(T\right)$, and density distributions,
$p\left(\rho\right)$ for the TO simulation with two different $\feh$
evaluated on layers of constant Rosseland optical depth, in order
to illustrate the differences in the statistical properties. The histogram
of the metal-poor case differs substantially in upper layers from
the solar one. Furthermore, in Fig. \ref{fig:tt_hist}, we show $p\left(T\right)$
and $p\left(\rho\right)$ in the upper layers ($\left\langle \ltaur\right\rangle =-4.0$)
for dwarf models with different $\teff$ and $\feh$. In both cases
we compare the distributions on constant geometrical height $z$,
constant column mass density $m$ and constant Rosseland optical depth
$\taur$.

At solar metallicity (Fig. \ref{fig:tt_hist}), the temperature distributions
are very narrow and symmetric. With increasing $\teff$, the average
$T$ is as expected higher and the width of the distribution broadens
slightly. The mean values are very similar between the different $\hav$
methods and in principle indistinguishable, which also agrees with
Fig. \ref{fig:temp}. Furthermore, the mean values are located very
close to the mode.

At $\feh=-3.0$, the temperature distributions change considerably.
While at cooler $\teff$ the shape is vey narrow and symmetric, for
$\teff\ge5500\,\mathrm{K}$ we find a distinct broadening of the $T$-distribution
on geometrical reference depth scale $\havz$, which is given by a
long tail at high $T$ and a decreasing peak at lower $T$ (see Figs.
\ref{fig:hist_to_ov} and \ref{fig:tt_hist}). In the column mass
density averages $\havcm$ the temperature peak is slightly more pronounced
at higher $\teff$, while the high-$T$ tail is slightly reduced.
The situation is pretty different for the averages on Rosseland optical
depth $\havr$, where we find that the temperature peak drops faster
towards higher $\teff$, and at $7000\,\mathrm{K}$ the $T$-distribution
shows an almost unimodal distribution. The mean values disagree at
higher $\teff$ between the different reference depth scales.

The density distributions behave differently depending on the reference
depth scale. On $\havz$ the histograms are in general slightly skewed
with a fat tail towards lower $\rho$ for all metallicities (Figs.
\ref{fig:hist_to_ov} and \ref{fig:rho_hist}). The density distributions
for the averages on column mass density are very symmetric and narrow
for both solar and low metallicities. At solar metallicity, the density
histograms on constant optical depth are narrower and higher than
the geometrical analogs, but skewed in contrast to $\havcm$. In the
metal-poor case, $\left\langle p\left(\rho\right)\right\rangle _{\mathrm{Ross}}$
becomes very narrow and symmetric at lower $\teff$, but towards higher
$\teff$ we find the $\rho$-distribution to also be broader. The
mean density stratification varies considerably among the different
averaging methods.

As mentioned above, adiabatic cooling due to mechanical expansion
and radiative reheating are competing with each other in the upper
photosphere and contribute to the phenomenon of reversed granulation.
At lower metallicity, the reversed granulation is enhanced, so that
the optical depth is increasingly strongly corrugated towards higher
$\teff$, which in turn will amplify the differences in statistical
properties during the translation to the optical depth scale from
the geometrical depth scale. This leads to the systematical broadening
in the statistical distribution that we encounter at lower metallicity.

\section{Spectral line formation: $\hav$ and $3\mathrm{D}$ LTE calculations\label{sec:Spectral-line-formation}}

\begin{figure*}[t]
\includegraphics[width=176mm]{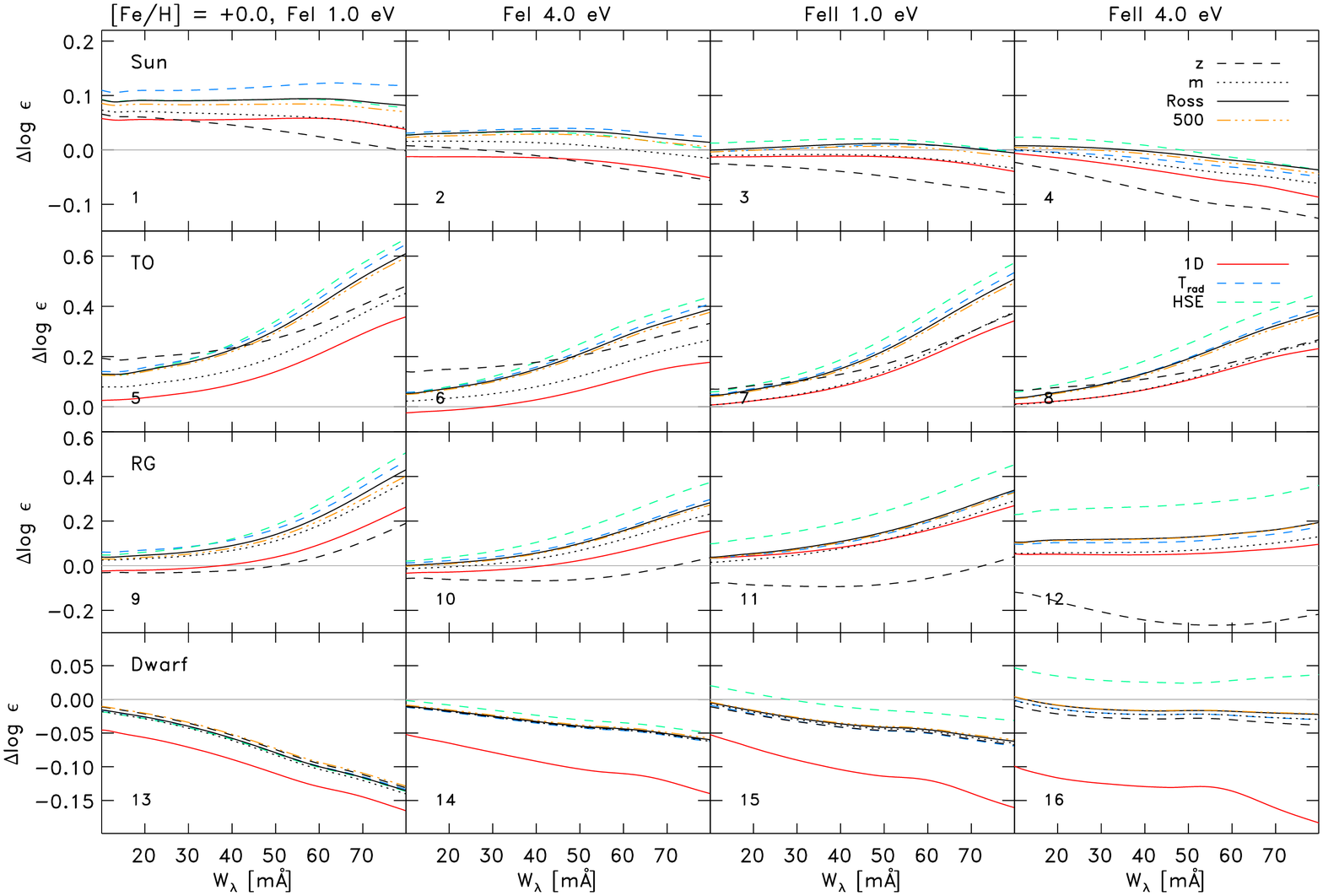}

\caption{Overview of the $\hav-3\mathrm{D}$ line formation differences given
in abundances displacement $\dlga$ vs. equivalent width $\eqw$ for
the $\fei$ and $\feii$ fictitious spectral lines with the excitation
potentials $\xex=1.0$ and $4.0\,\mathrm{eV}$ including the Sun,
TO, RG and dwarf simulation (from top to bottom). The averages on
layers of constant geometric height $\havz$ (black dashed), constant
column mass density $\havcm$ (black dotted), constant Rosseland optical
depth $\havr$ (black solid) and at 500 nm $\havf$ (orange dashed
triple-dotted lines) are indicated. Furthermore, we show 1D models
(red solid), $T_{\mathrm{rad}}^{\mathrm{Ross}}$-averages (blue dashed)
and $\havr^{\mathrm{HSE}}$ (green dashed lines). The microturbulence
of $\mturb=1.0\,\mathrm{km}/\mathrm{s}$ has been used throughout.
Notice the different ordinates.}

\label{fig:fakelines1} 
\end{figure*}

\begin{figure*}[t]
\includegraphics[width=176mm]{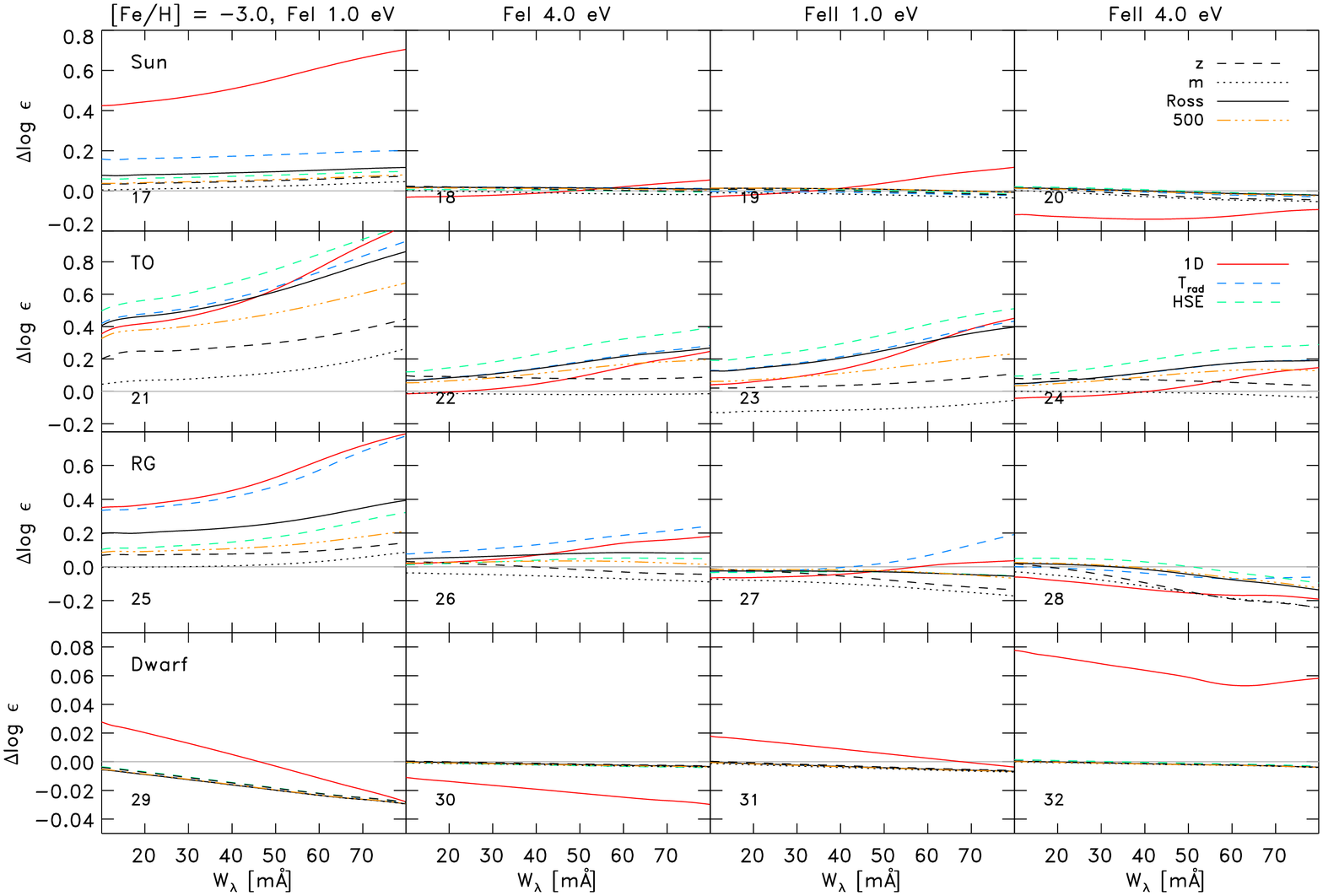} \caption{Similar to Fig. \ref{fig:fakelines1} but showing overview of the
abundance corrections for metal-poor models, with larger ranges for
the $y$-scales.}

\label{fig:fakelines2} 
\end{figure*}

To explore the differences between the line formation based on $\hav$
and full 3D models, we have chosen a set of representative models
consisting of a main-sequence (MS) star ($\teff/\logg=5777$~K/$4.44$),
a turn-off (TO) star ($6500$/$4.0$), a red-giant (RG) star ($4500$/$2.0$),
and a dwarf ($4500/5.0$). For all these models, we considered metal-poor
analogs with $\feh=-3.0$ besides the solar metallicity.

\subsection{3D line formation calculations\label{sub:SCATE}}

We used the 3D radiative transfer code \textsc{Scate} \citep{Hayek:2011p8560}
to calculate full 3D synthetic spectral line disk-center intensity
and flux profiles with 3D \textsc{Stagger} model atmospheres. \textsc{Scate}
assumes local thermodynamic equilibrium (LTE). Furthermore, in the
present work, we also neglected the effects of scattering; i.e. we
approximated the source function with the Planck function, $S_{\lambda}=B_{\lambda}$.
We caution that LTE is in general a poor approximation, especially
for $\fei$ spectral line formation calculations at low $\feh$ \citep[e.g.][]{Bergemann:2012p20128},
which should be kept in mind for analyzing the LTE-based abundance
corrections presented here. For the sake of consistency, we used the
same EOS \citep{Mihalas:1988p20892} and continuum opacity data \citep[from the MARCS package; see][]{Gustafsson:2008p3814}
as in the 3D \textsc{Stagger} simulations.

To reduce the computational costs for line formation calculations,
we consider a subset of $N_{t}=20$ temporally equidistant snapshots
-- the same as used for the temporal $\hav$ averages -- sampling
the entire time spans of the individual 3D simulation sequences. Additionally,
we reduce the horizontal spatial resolution from $N_{x}N_{y}=240^{2}$
to $60^{2}$ by considering only every fourth column in each horizontal
direction. Test calculations carried out at full resolution show that
differences are negligible for all practical purposes \citep[see][]{Asplund:2000p20875}.
Concerning the vertical direction, while we did not subsample the
number of depth points, we considered only those layers with $\min(\ltaur){\leq}3.0$.
The resulting disk-center intensity and flux profiles are spatially
and temporally averaged, and then normalized with the respective continuum
intensity or flux.

To systematically illustrate the differences between $\hav$ and 3D
line formation, we computed \textsl{\emph{fictitious}} atomic lines
for neutral and singly ionized iron, $\fei$ and $\feii$, for the
selected \textsc{Stagger}-grid models and metallicities. All lines
are defined at the same wavelength, $\lambda=500\,\mathrm{nm}$, and
we considered two lower-level excitation potentials, $\xex=1.0$ and
$4.0\,\mathrm{eV}$. Furthermore, we varied the oscillator strength,
$\log gf$, in order to cover a range of line strengths, from weak
to partly saturated lines, with equivalent widths from $\eqw=5$ to
$80\,\mathrm{m\angs}$. We assumed an iron abundance of $\log\epsilon_{\mathrm{Fe}}=7.51$
\citep{Asplund:2009p3308} and $\log\epsilon_{\mathrm{Fe}}=4.51$,
for the solar metallicity and $\feh=-3.0$ case, respectively.

The spectral line calculations with $\hav$ models were also performed
with \textsc{Scate}, to guarantee a consistent comparison. \textsc{Scate}
employs atmospheric structures on geometrical height and computes
the optical depth, $\tau_{\lambda}$, for the individual line. Therefore,
we provide the geometrical height by integrating $dz=d\left\langle \tau_{\lambda}\right\rangle /\left\langle \kappa_{\lambda}\right\rangle $,
which is of course unnecessary for $\havz$. Furthermore, tests revealed
that including just an averaged velocity, e.g. $\left|\vec{v}\right|/3$,
is insufficient to reproduce the influence of the 3D velocity field
on the line shape. Analyzing the influence of the velocity field on
the line formation surpasses the scope of the present work; therefore,
we will explore this aspect in a separate study. In this paper, for
the calculations with $\hav$ models we neglected the information
about the actual velocity field and instead assumed a fixed microturbulence
of $\mturb=1.0\,\mathrm{km}/\mathrm{s}$ for all considered stellar
parameters.

Since the line formation calculations with $\hav$ models are obviously
much faster, we use the $\havr$ averages first to estimate the $\lgf$
range, which would result in the designated range in $\eqw$. We then
consider ten equidistant $\lgf$ values within that range for the
$\hav$ and full 3D models. Finally, we interpolate the curves of
growth ($\lgf$ vs. $\eqw$) using a spline interpolation and retrieve
%% from the latter
the $\Delta\lgf$ difference between $\hav$ and $3\mathrm{D}$ synthetic
lines at a given equivalent width; i.e., $\Delta\lgf=\hav-3\mathrm{D}$.
For trace elements, changes in line strength due to $\Delta\lgf$
are equivalent to changes due to abundance variations $\dlga$; hence,
the $\Delta\lgf$ differences can be interpreted as $\hav-3\mathrm{D}$
abundance corrections. With four fictitious lines and four representative
models with two metallicities, we covered 32 cases in total.

\label{sub:Consequences-line-formation}Full 3D line profiles are
marked by line shifts and asymmetries owing to the non-thermal Doppler
broadening introduced by the up- and downflows of the convective motions,
which are present in the photosphere due to overshooting \citep{Asplund:2000p20875}.
In 3D RHD modeling, the velocity field emerges naturally from first
principles. The buoyant hot rising plasma in the granules blue-shifts
the line, while the fast downdrafts introduce a redshift. Besides
the convective motions, another source of line broadening are the
inhomogeneities in the thermodynamic independent variables, $\rho$
and $T$. The ascending granules are hotter and less dense than the
downdrafts (see Fig. \ref{fig:tt_updn}). The velocities and inhomogeneities
prevailing at formation height of the individual lines will lead to
line shifts and asymmetries. The $\hav$-based lines are symmetric
without any shifts, however, we can compare the equivalent widths
of lines from calculations based on full 3D models and on the different
average stratifications.

We probed different formation heights with the parameters of our fictitious
lines. The $\feii$ lines form deeper in the atmosphere, closer to
the continuum forming layers, while the $\fei$ lines are more sensitive
to the intermediate heights of the atmosphere. Spectral lines with
lower (higher) excitation potential form at smaller (larger) optical
depths. We showed in Sect. \ref{sec:Comparison-of-averages} that
the metal-poor model stellar atmospheres exhibit rather different
temperature stratification at the top depending on the averaging method,
consequently the latter should show the largest differences between
the $\hav$ models.

\subsection{Comparison of $\hav$ and $3\mathrm{D}$ line formation\label{sub:Comparison-line-formation}}

We show an overview of the differences between the $\hav$ and the
full 3D calculations in Figs. \ref{fig:fakelines1} and \ref{fig:fakelines2}.
The first noticeable observations are the systematic trends in form
of a slope towards higher line strength, which are due to the fixed
value of the microturbulence, $\mturb$, with $1\,\mathrm{km}/\mathrm{s}$
in the $\hav$ models. An increasing slope with line strength indicates
an underestimation of $\mturb$, in particular for the TO and RG (see
panel 5 to 12 in Fig. \ref{fig:fakelines1} and 21 to 28 in Fig. \ref{fig:fakelines2}).
By contrast, in cool dwarfs, the adopted $\mturb$ seem to be overestimated.
These findings agree with comparisons of 1D models with observations
\citep[e.g., ][]{Edvardsson:1993A&Ap275,Bensby:2009A&Ap499}. We tested
this by applying a number of $\mturb$ values%
\footnote{We find a reduction of the slope in the curve-of-growth with $\mturb=0.5,\,1.5,\,2.0\mathrm{km}/\mathrm{s}$
for the dwarf, RG and TO models respectively (while a fine-tuning
could flatten it completely). %
}, which showed that a fine-tuning can rectify the present slope. However,
for the sake of clarity, we prefer to limit the already large number
of stellar and line parameters to just a single $\mturb$. The calibration
of the microturbulence will be the subject of a separate study.

Weak lines are insensitive to $\mturb$, yet they show variations
in strength, which can be attributed to differences in the mean $\hav$
stratifications of temperature and density. Interestingly, when one
compares this regime between the different averages in Fig. \ref{fig:fakelines1},
the averages on column mass density are often the closest to the full
3D spectral lines and perform in this respect often better than the
averages on constant Rosseland optical depth. The stratification on
constant optical depth at 500 nm always shows spectral line features
slightly closer to the full 3D case compared to the Rosseland optical
depth. However, this is because we chose our fictitious iron lines
at $500\,\mathrm{nm}$, which leads to an inherent advantage of $\havf$
over $\havr$. The geometrical averages show large deviations in the
case of the TO and RG star at solar metallicity (see panels 5 to 12).

The differences in the metal-poor case (Fig. \ref{fig:fakelines2})
are clearly greater than in the solar metallicity models (Fig. \ref{fig:fakelines1}).
It is obvious that $\hav$ models at low $\feh$ struggle to reproduce
the 3D case properly, in particular $\fei$ lines with small excitation
potential, and the differences are particularly pronounced for the
hotter metal-poor TO stars (panel 21). This is in accordance with
our findings from Sects. \ref{sec:Comparison-of-averages} and \ref{sec:Statistical-properties}:
at low metallicity and high $\teff$. The differences in the statistical
properties among the various $\hav$ averages increases at low $\feh$.
In particular, the widths of the temperature and density distributions
become broader at lower metallicity (Fig. \ref{fig:tt_hist}), and
their mean values become increasingly less well-defined in its statistical
representation. The reason for the broadening is the enhanced contrast
of the reversed granulation due to the reduced radiative re-heating
with weak spectral line features at low metallicity (see App.\ref{app:Reversed-granulation}).

\begin{figure*}
\includegraphics[width=88mm]{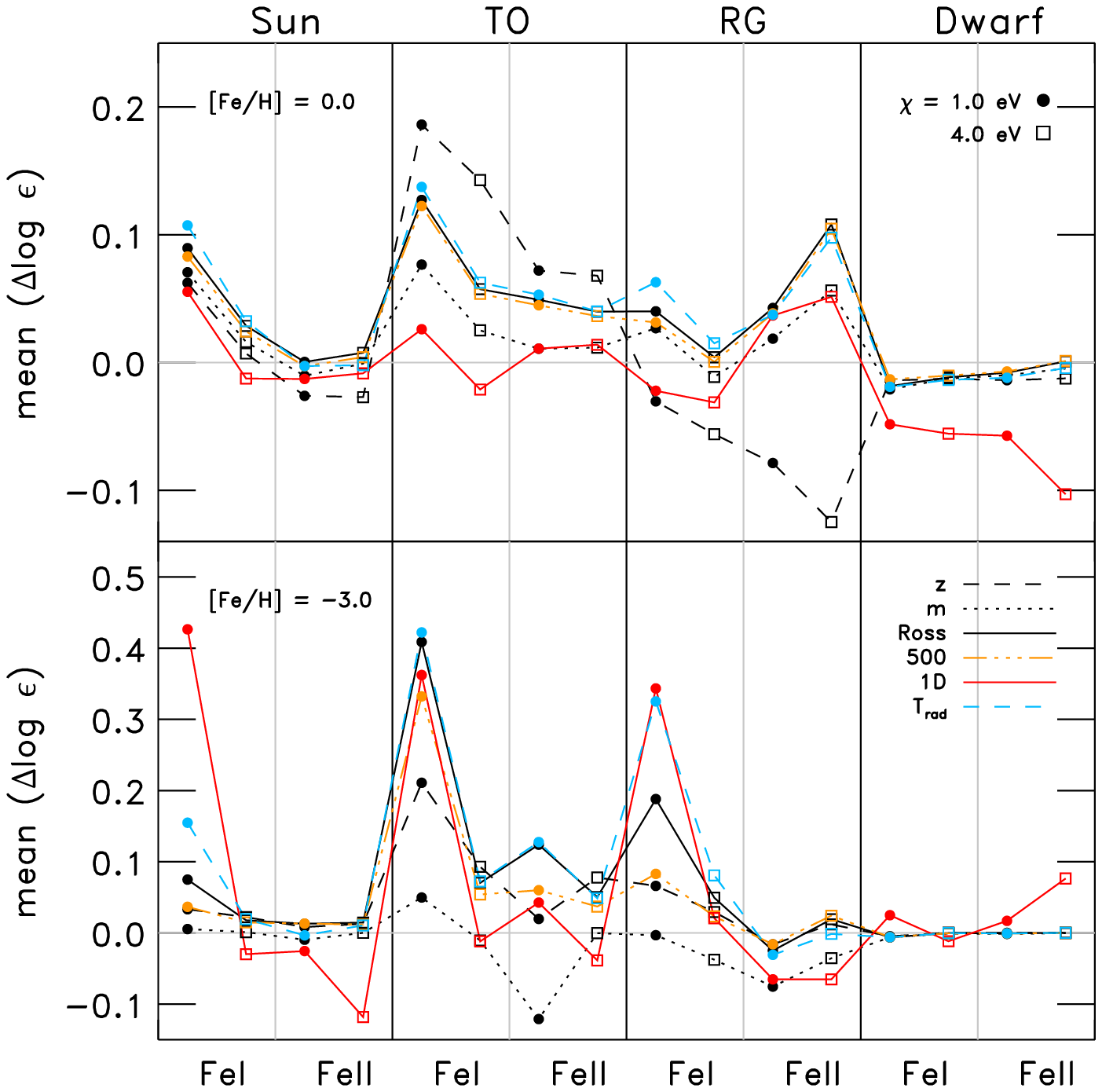}\includegraphics[width=88mm]{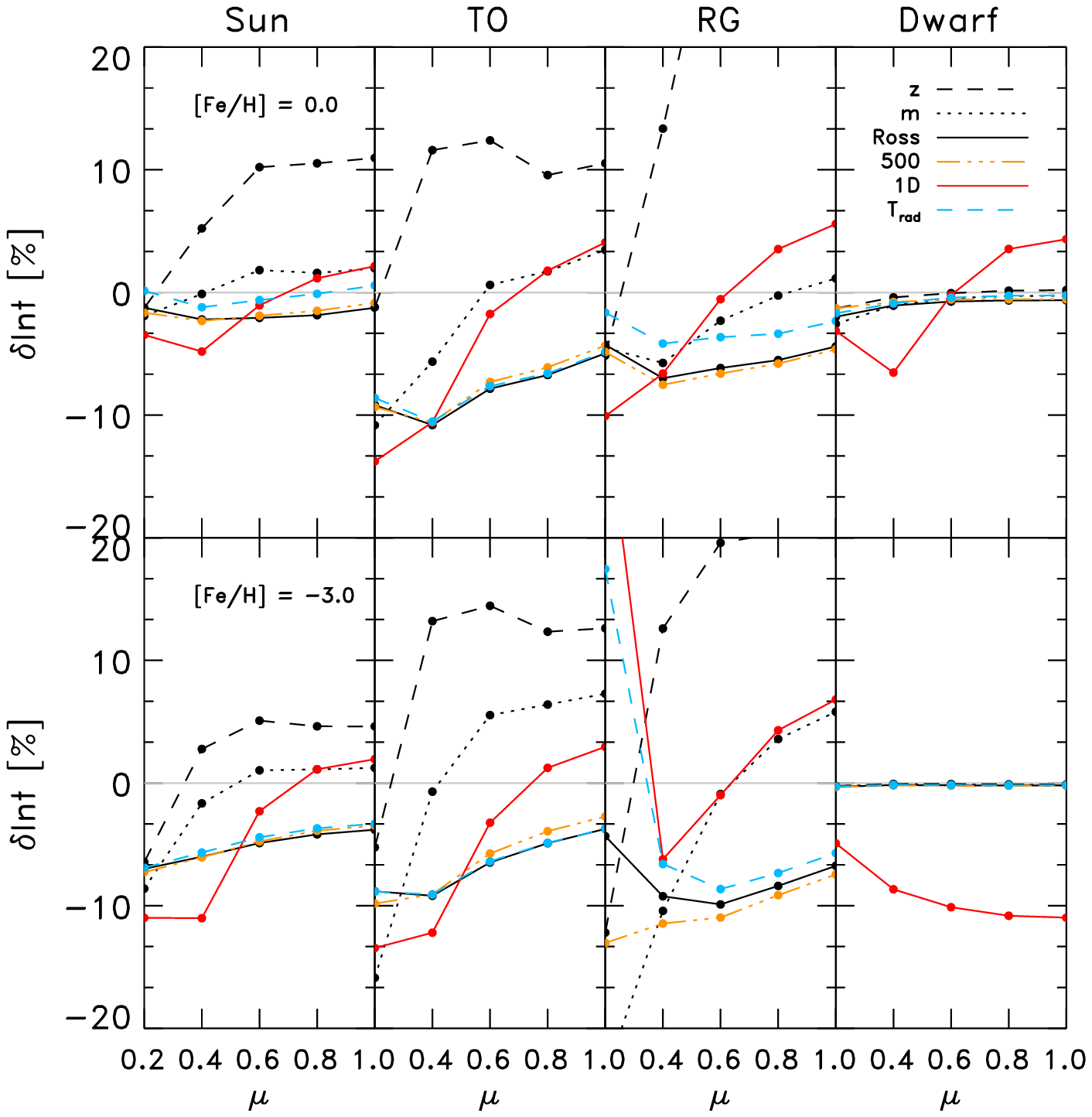}

\caption{In the left figure the mean $\Delta\log\varepsilon$ (evaluated between
$5-20\,\mathrm{m}\angs$) is illustrated against $\fei$ and $\feii$
given at $\xex=1.0$ and $4.0\,\mathrm{eV}$ for the different selected
models. In the right figure, the relative difference with $\hav-3\mathrm{D}$
of the continuum intensity, $\delta I_{\mu}$, vs. $\mu$ angle is
displayed. Both Figures include the solar metallicity (top) and the
metal-poor (bottom) case, and the averages $\havz$ (black dashed),
$\havcm$ (black dotted), $\havr$ (black solid), $\havf$ (orange
dashed triple-dotted), $T_{\mathrm{rad}}^{\mathrm{Ross}}$-averages
(blue dashed), and 1D models (red solid lines).}

\label{fig:ov_fakelines}\label{fig:ov_clv} 
\end{figure*}

To facilitate an overall comparison between the different averages
with respect to line formation, we show in Fig. \ref{fig:ov_fakelines}
(left) the mean abundance deviations for weak lines that are determined
between $\eqw=5-20\,\mathrm{m}\angs$. For the model representing
the Sun, the differences between $\hav$ and 3D are in general small:
$\lesssim0.1\,\mathrm{dex}$. For the TO stars at solar $\feh$, the
differences are considerably larger: $\lesssim0.2\,\mathrm{dex}$.
We find the largest deviations for $\fei$ lines with small excitation
potential $\xex=1.0\,\mathrm{eV}$, which are the most temperature
sensitive; in particular the geometrical averages exhibit strong differences.
At lower metallicity, the differences increase in particular for the
TO and RG model with $\lesssim0.4\,\mathrm{dex}$, and the $\hav$
on optical depth shows the largest deviation for metal-poor TO star.
In general the deviations become smaller at higher $\xex$ and for
$\feii$ lines. The dwarfs show very small differences compared to
the full 3D case. These models exhibit the lowest velocities and temperature
contrast with the mean stratifications closely resembling the 1D models
based on same EOS and opacities.

The averages on column mass density $\havcm$ typically exhibit the
best agreement with the predictions of the full 3D model, in particular
at low metallicity. The geometrical averages $\havz$ exhibit large
deviations \citep[in agreement with ][]{Uitenbroek:2011p10448}, especially
for the TO stars. When one considers the comparison of the temperature
and density in Fig.~\ref{fig:temp}, then one can deduce that the
models with cooler stratifications are closer to the full 3D line
strength. Both models averaged on constant optical depth, $\havr$
and $\havf$, lead to systematically larger deviations from the full
3D line formation calculations than those obtained with $\havcm$
models, in particular for low excitation $\fei$ for the metal-poor
TO star.

The resulting spectral line features with the logarithmic averages
$\hav_{\log}$ are similar to plain $\havr$ (therefore we refrain
from showing the latter), while averages enforcing hydrostatic equilibrium,
$\havhse$, clearly fail to closely reproduce the results from 3D
line formation \citep[similar to][]{Uitenbroek:2011p10448} and lead
to rather large errors in the line formation, in particular for the
metal-poor TO model (Fig. \ref{fig:fakelines2}). Furthermore, both
the flux-weighted and brightness-temperature averages, $T^{4}$ and
$T_{\mathrm{rad}}$, are in general very close to the plain average,
but often slightly less accurate, which is a somewhat surprising result
(see $T_{\mathrm{rad}}$ in Fig. \ref{fig:ov_fakelines}). 

Another meaningful way to test the performance of the different averages
can be accomplished by comparing the deviation of the center-to-limb
variation (CLV) of the continuum intensity. In Fig. \ref{fig:ov_clv},
we show the differences of the continuum intensity, $\delta I_{\mu}=(I_{\mu}^{\hav}-I_{\mu}^{3\mathrm{D}})/I_{\mu}^{3\mathrm{D}}$,
i.e. between the $\hav$ and full 3D models. We find in general that
the $\hav$ models overestimate the continuum intensity at disk center
($\mu=1$), while towards the limb ($\mu=0.2$) the $\hav$ often
underestimate the intensity. The deviations of the different averages
are similar to the above findings with the comparison of the curve
of growth. The disk-center intensities of the 3D RHD models are matched
best by the averages on column mass density $\havcm$, whereas the
geometrical averages $\havz$ display the largest discrepancies, in
particular for the RG model at solar metallicity with an overestimation
by $\sim60\,\%$. The results for the averages on optical depth are
once again midway between the two other kinds of averages. An interesting
aspect is that the brightness-temperature averages $T_{\mathrm{rad}}$
fail to render the continuum intensities exactly, which has to be
interpreted as a consequence of the non-linearity of the Planck function.
Our findings are qualitatively similar to those by \citet{Uitenbroek:2011p10448}.

\subsection{Cautionary remarks\label{sub:Cautionary-remarks}}

We remind the reader that LTE is often a very poor assumption at low
$\feh$ \citep[e.g.][]{Asplund:2005p7792} and thus that the abundance
differences presented in Figs. \ref{fig:fakelines1} and \ref{fig:fakelines2}
should not be added indiscriminately to results from standard 1D LTE
abundance analyses. In LTE, the difference between 3D and 1D models
can be very substantial for metal-poor stars for especially low excitation
and minority species like $\fei$ \citep[e.g.,][]{Asplund:1999p11771,Collet:2007p5617},
but those same lines also tend to be sensitive to departures from
LTE \citep[e.g.,][]{Bergemann:2012p20128,Lind:2012p427} in 1D and
$\hav$ models, mainly due to overionization and overexcitation in
the presence of a hotter radiation field than the local kinetic temperature
(i.e., $J_{\lambda}>B_{\lambda}$). Although not explored for more
than Li, one would expect that the very cool upper atmospheric layers,
hence steep temperature gradients in metal-poor 3D models compared
with classical 1D models, are even more prone to substantial non-LTE
effects \citep[e.g.,][]{Asplund:2003p7793,Sbordone:2010p10214}. In
particular, neutral species of relatively low ionization energy, such
as $\fei$, typically suffer from significant positive NLTE abundance
corrections due to overionization \citep[e.g.,][]{Asplund:2005p7792,Bergemann:2012p20128,Lind:2012p427}
with low excitation lines are especially prone. For low-excitation
$\fei$ lines, one would therefore expect the 3D NLTE line strengths
to be more similar to the 1D case than the 3D LTE results due to the
positive NLTE corrections, partly compensating for the negative 3D
LTE corrections. We therefore caution the reader that the 3D LTE abundance
corrections presented here (3D LTE - 1D LTE) for $\fei$ lines are
likely to be too negative compared to the NLTE case (3D NLTE - 1D
NLTE). As a corollary, it is inappropriate to apply a 1D NLTE abundance
correction to a 3D LTE-inferred abundance when the latter is very
significant, as is often the case at low $\feh$.

\subsection{Comparison with 1D models\label{sub:1D-models}}

In Paper I we compared the $\havr$ stratifications with 1D models
computed with the same EOS and opacity as used in the \textsc{Stagger}-code,
in order to quantify the differences arising solely from 1D modeling
based on MLT. The line formation calculations with 1D models perform
quite well at solar metallicity, with the exception of the cool dwarf
models (Fig. \ref{fig:fakelines1}). However, in the metal-poor case,
the lines based on the 1D models obviously do not correctly reproduce
the full 3D lines by overestimating the $T$-stratifications due to
the enforcement of radiative equilibrium in the upper atmosphere (Fig.
\ref{fig:fakelines2}). This is, in particular, distinctive for low-excitation
neutral iron lines as previously found by \citet{Asplund:1999p11771}
and \citet{Collet:2007p5617}. \citet{Kucinskas:2012p23943} present
similar findings for a solar-metallicity RG simulation as well, namely
that neutral iron lines based on 1D MLT models are slightly closer
to the full 3D lines compared to the $\hav$ lines.

We note that in our 1D models the turbulent pressure is neglected,
and the mixing length is fixed with $\alpha_{\mathrm{MLT}}=1.5$,
both choices that will influence the stratification significantly.
Since their effect is strongest in convective zone below the optical
surface and the line formation region, the influence in terms of abundance
is likely small; in fact, \citet{Kucinskas:2012p23943} only found
a very small effect $<0.02\,\mathrm{dex}$ for the reduction in $\alpha_{\mathrm{MLT}}$
from 1.5 to 1.2. However, for metal-poor giants the influence can
be greater for lines with very high excitation potential.

\section{Conclusions\label{sec:Conclusions}}

We have investigated the properties of different methods in detail
for computing temporal and horizontal average stratifications from
3D RHD \textsc{Stagger}-grid simulations of stellar surface convection.
The choice of the reference depth is critical, as comparisons of the
various $\hav$ demonstrated. We find in general that the temperature
stratifications of the $\havz$ and $\havcm$ are hotter close to
the continuum forming layers and cooler in the upper layers compared
to averages on surfaces of constant optical $\havr$ and $\havf$,
while the density shows differences in the opposite sense. The flux-weighted
temperature average and brightness temperature average are distinctively
hotter than the plain averages, both close to the optical surface
and in the upper atmosphere, since the Planck function and the fourth
powers weights the larger temperatures higher. Averages obtained from
the logarithmic values lead to lower temperature and density distributions
by giving more weight the lower values in the distribution. These
characteristics increase with higher $\teff$, lower $\logg$ and
especially with lower $\feh$.

The statistical properties change depending on the reference depth
scale, since the transformation to the new depth scale will inevitably
imply a remapping of the values from different heights. The translation
to layers of constant optical depth will smooth out temperature fluctuations
as a byproduct: the temperature is in fact the main source of spatial
corrugation of the surfaces of constant optical depth due to the strong
temperature sensitivity of the dominant $\mathrm{H}^{-}$ continuum
opacity source. Therefore, the temperature contrast and extrema are
distinctively reduced, in particular in the superadiabatic region.
However, this has also the side effect of enhancing both contrast
and minimum-maximum range of the density. The concomitant remapping
of properties from deeper or higher layers during the transformation
to the new reference depth scale will in turn change the average values.

Furthermore, we examined the effects of reversed granulation in the
upper layers of metal-poor stars, namely the lowering of temperatures
above the granules in metal-poor 3D models compared to classical 1D
models. We found that the contribution of radiative reheating due
to weak spectral line absorption features relative to cooling due
to mechanical expansion in the upper atmospheric layers is reduced
towards higher $\teff$. On the other hand, the temperature in the
regions immediately above the intergranular lanes are primarily controlled
by mechanical expansion or compression and do not appear to be affected
by the reduced metallicity. The two combined effects result in an
enhanced contrast in the reversed granulation. This in turn leads
to an increase in the corrugation of the surfaces of constant optical
depth, which implies that the averages on constant optical depth are
sampling values from a very wide range in geometrical height, thereby
affecting the statistical properties such as mean value and contrast.

The comparison of $\fei$ and $\feii$ calculated in full 3D and different
$\hav$ atmosphere models reveals the surprising result that the averages
on column mass density $\havcm$ typically provide the best representation
of the 3D model with respect to the line formation. The commonly preferred
averages on layers of constant optical depth $\havr$ or $\havf$
in general perform worse. We located the reason for the underperformance
in the predictions of 3D RHD by the $\hav_{\tau}$ models being due
to the optical depth, $d\tau_{\lambda}=\rho\kappa_{\lambda}dz$, which
contains the additional non-linearity of opacity $\kappa_{\lambda}$,
in contrast to the column mass density, $dm=\rho dz$; therefore,
the statistical properties, in particular, the mean value, are more
prone to distinctive temperature fluctuations present in the superadiabatic
region and the upper layers, where the reversed granulation takes
place. The differences between the lines calculated with the $\hav_{\tau}$
models and the full 3D RHD models are significant, in particular,
for metal-poor simulations due to the enhanced reversed granulation
in the upper layers. We find that the neutral $\fei$ lines with low
excitation potential feature the largest differences between the mean
$\hav$ and full 3D line calculations. The 1D MLT models perform quite
well at solar metallicity; however, for metal-poor models the mismatch
is evident. Therefore, we caution against using 1D models for metal-poor
stars, which will lead to systematic errors in the spectral analysis.

For spectral line formation calculations with $\hav$ models from
the \textsc{Stagger}-grid, we recommend using averages obtained on
layers of constant column mass density, $\havcm$, since these provide
the closest match to the spectral line strengths obtained with the
full 3D RHD models. Furthermore, we advise strongly against using
geometrical averages $\havz$ for spectral line formation calculations.
For purposes of improving stellar structures and asteroseismology,
the $\havz$ models are, however, useful, since these averages alone
fulfill the hydrostatic equilibrium, and therefore, comparisons with
helioseismological observations are in better agreement.

It is obvious that the temporally and spatially averaged models are
incapable of substituting the full 3D atmospheric structure. The reduction
due to the averaging will unavoidably lead to sacrificing required
information. A promising intermediate approach could be the so-called
\textquotedbl{}1.5D\textquotedbl{} approximation. This approach emulates
atmospheric inhomogeneities, which are probed by the traversing radiation,
by considering a series of perturbed 1D stratifications for spectral
synthesis \citep[e.g., see][]{Ayres:2006p21937}. In the spirit of
the latter, one could utilize the temporal averaged histograms for
an improved spectral line synthesis, since these contain additional
information on the statistical distribution of the 3D simulations.
\begin{acknowledgements}
We acknowledge access to computing facilities at the Rechenzentrum
Garching (RZG) of the Max Planck Society and at the Australian National
Computational Infrastructure (NCI) where the simulations were carried
out. Remo Collet is the recipient of an Australian Research Council
Discovery Early Career Researcher Award (project number DE120102940).
We thank Tiago Pereira for his contribution.
\end{acknowledgements}
\bibliographystyle{aa}
\bibliography{papers}

\appendix
%dummy comment inserted by tex2lyx to ensure that this paragraph is not empty

\section{Addendum to averaged models\label{app:Remarks-on-averages}}

\subsection{Reversed granulation\label{app:Reversed-granulation}}

\begin{figure}
\includegraphics[width=88mm]{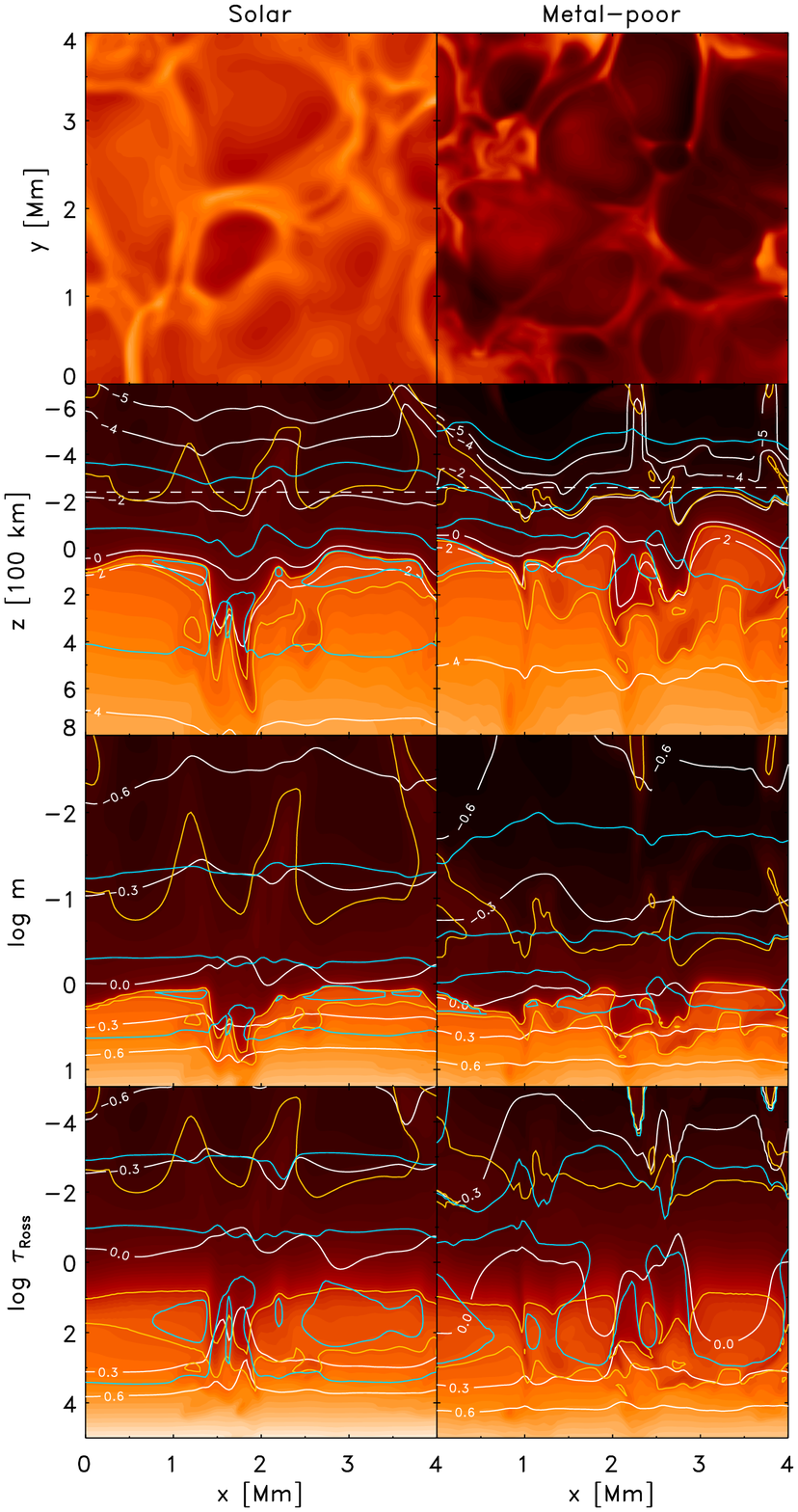} \caption{Temperature contours from our model with $\teff=6500\,\mathrm{K}$
and $\logg=4.5$ with $\feh=0.0$ (left) and $-3.0$ (right). The
top panels display horizontal slices with the reversed granulation
pattern imprinted in the temperature map (from 3 to $7\times10^{3}\,\mathrm{K}$)
taken at $\sim230\,\mathrm{km}$ above the surface, which is also
indicated in the second panel (dashed lines). The panels below show
vertical slices ($T$-contours from 2 to $17\times10^{3}\,\mathrm{K}$)
ranging from $-5.0\le\ltaur\le5.0$ on layers of constant geometrical
height (second), column mass density (third) and Rosseland optical
depth (last panel). These panels include isocontours of the temperature
($5,\,10$ and $12\times10^{3}\,\mathrm{K}$; yellow lines) and density
($0.1,\,1.0$ and $2.5\times10^{-7}\mathrm{g}/\mathrm{cm}^{3}$; blue
lines) and both increase with decreasing vertical depth. We show also
lines of constant optical depth (second) and geometrical depth (third
and last) indicated with white lines.}

\label{fig:reversed_granulation} 
\end{figure}
To illustrate the effects of the remapping of the 3D atmospheric structures
on new reference depth scales, we show slices of temperature contours
from our TO-simulation in Fig. \ref{fig:reversed_granulation}. We
show horizontal temperature maps taken in the upper atmosphere (top
panel) and three vertical slices with different reference depth scales,
which include geometrical $z$ (second panel), column mass density
$m$ (third panel), and Rosseland optical depth (bottom panel). Furthermore,
we indicate three different isocontours of the temperature (yellow)
and density (blue lines) in Fig. \ref{fig:reversed_granulation},
and we also show lines of constant optical depth $\taur$ (white lines
in top panel) or geometrical depth $z$ (middle and bottom panels). 

The downdrafts just below the optical surface, which are denser and
cooler than the lighter and hotter surrounding granules, are easily
identified (by the prominent changes in $T,\rho$ and $\taur$ above
the downflows, e.g. $x\approx1.8\,\mathrm{Mm}$). Owing to the lower
temperatures in the downdrafts compared with the granules, the same
optical depth value is reached at lower geometrical depths, meaning
that the emergent radiation in the intergranular lanes originate in
much deeper geometrical heights. The corrugation of the optical depth
on geometrical depth scale is therefore most pronounced in the downdrafts
(see isocontour of $\ltaur=2.0$ in second panel of Fig. \ref{fig:reversed_granulation}).

The opposite is true for the upper atmospheric layers because of the
phenomenon of \emph{reversed granulation} \citep{Rutten:2004p16166,Cheung:2007p16147},
namely, above the granules, cooling by adiabatic expansion is dominant,
while above the inter granular lanes the radiative reheating and mechanical
compression are more important for the energy balance. At lower metallicity
and higher $\teff$, the radiative heating above granules is reduced
by the weakening of spectral line features. The resulting reduction
in radiative reheating leads to significantly cooler temperatures
(see top panel in \ref{fig:reversed_granulation}) and a lower pressure
support, and as a consequence the atmospheric layers at a given constant
optical depth subside toward lower geometrical heights, closer to
the optical surface. Therefore, the temperature contrast is enhanced
in the upper atmosphere. The subsiding of the atmosphere is similar
to what we found earlier, namely that the density range spanned in
the atmosphere is significantly reduced at lower metallicity (see
Fig. 16 in Paper I). Finally, the \emph{enhancement} of the reversed
granulation and the temperature contrast results in strongly corrugated
surfaces of constant optical depth at the top of metal-poor simulations.
We note that we also found an \emph{enhanced} intensity-contrast for
metal-poor stars (see Paper I).

The remapping of the individual columns of the 3D structure from geometrical
depth to optical depth entails a change of perspective between the
old and the new scales in terms of the distribution of values of a
particular physical variable at a given constant reference depth.
This is again most obvious in the downdrafts in the convection zone
(see line of constant geometrical depth at $z=0.2\,\mathrm{Mm}$ in
bottom panel of Fig. \ref{fig:reversed_granulation}). Properties
from deeper geometrical heights are mapped onto layers at lower optical
depth, and the temperature differences between upflowing and downflowing
regions are reduced, which results in a less of a temperature contrast
and in minimum-maximum ranges (see Sect. \ref{sub:Contrast}). On
the other hand, the deviations in the density are significantly \emph{enhanced},
which will clearly alter the statistical properties, in particular
the mean values.

In the upper atmospheric layers of the solar metallicity case, the
optical depth is corrugated only a small amount, therefore the transformation
does not affect the temperature and density much (compare the upper
flat blue line with the two lower corrugated ones in the bottom panel
of Fig. \ref{fig:reversed_granulation}). However, the corrugation
of the optical depth in the upper atmosphere is rather large for hotter
metal-poor stars owing to the enhanced reversed granulation. As a
result, the effects of remapping on the optical depth scale for the
temperature and density is fairly substantial in these simulations.
And the distribution of the thermodynamic properties is broadened,
such that the meaning of the horizontal average is weakened (see Fig.
\ref{fig:tt_hist}).

In a similar way, the translation to column mass density naturally
reduces the variations in density thanks to its definition of the
reference depth scale, which is the depth-integrated density. Therefore,
the resulting density fluctuations are rather small in layers at constant
column mass density. The variation in temperature is slightly lower
than in the averages on geometrical depth, but larger than in the
averages on optical depth, as one would expect.

We stress once again that the different reference depth scales are
equivalent to each other in terms of the spatial remapping of the
3D atmospheric structures. What differs of course is the statistical
properties of physical variables on layers of constant depth, which
vary depending on the choice of reference depth scale. One has to
consider two important aspects concerning the horizontal averaging,
the first being what kind of quantity is considered, and the second
which reference depth scale is accounted for. Therefore, the statistical
properties of the density and temperature are relatively distinctive
depending on which reference depth scale is considered (see Sect.
\ref{sec:Statistical-properties}).

\subsection{Hydrostatic equilibrium\label{app:hse_stratification}}

\begin{figure}
\includegraphics[width=88mm]{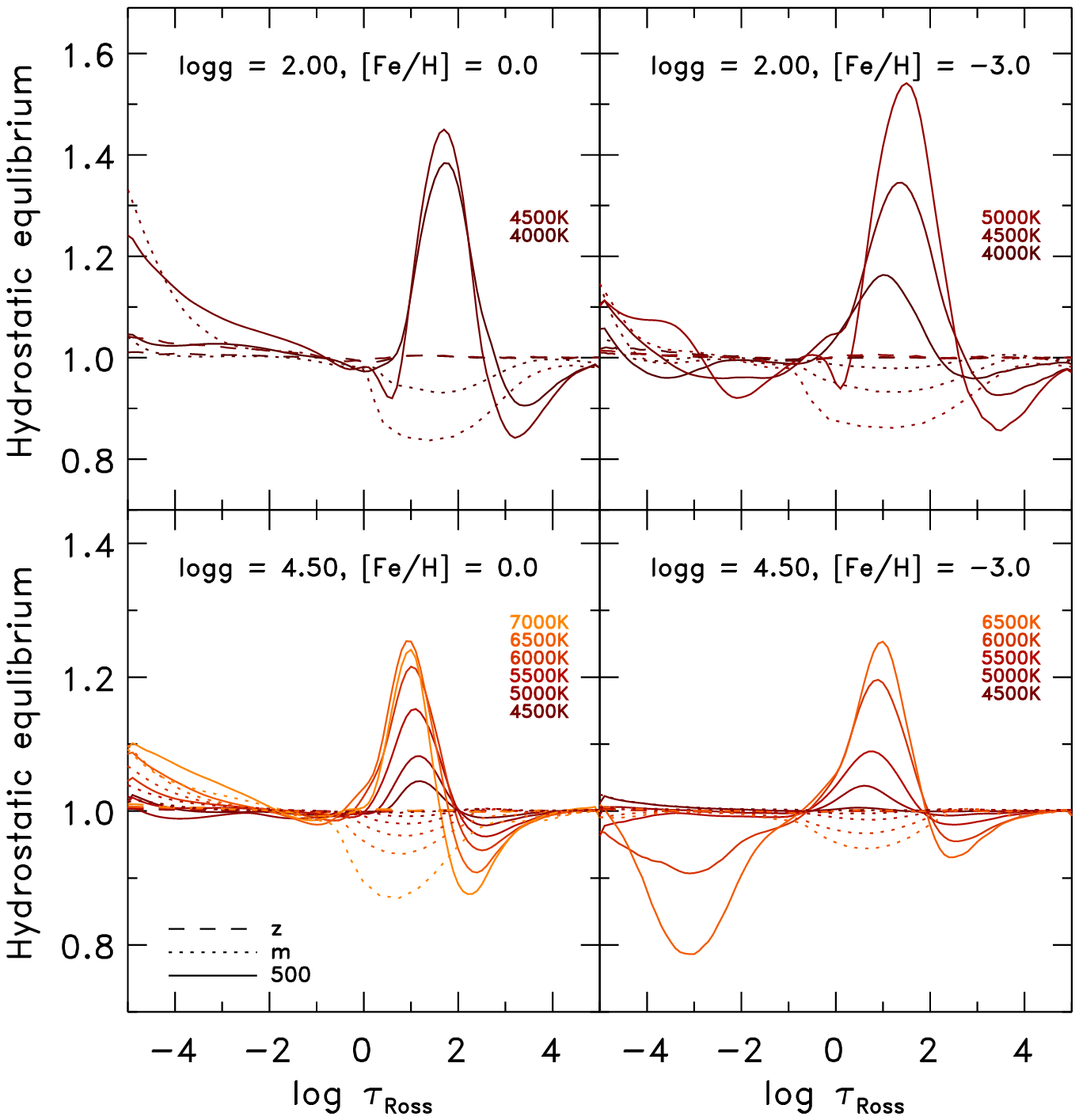}

\caption{Deviations from the hydrostatic equilibrium vs. optical depth. \emph{Dashed
lines}: $\havz$ averages; \emph{dotted lines}:$\havcm$; \emph{solid
lines}: $\havr$.}

\label{fig:hse} 
\end{figure}
The \textsc{Stagger}-code directly solves the discretized time-dependent,
radiative-hydrodynamical equations (see Paper I) for the conservation
of mass, momentum, and energy. The conservation properties are reflected
in the mean $\havz$ stratifications of relaxed, quasi-stationary
3D hydrodynamical models averaged on layers of constant geometrical
depth. In particular, the geometrical averages appear over time to
be close to hydrostatic equilibrium.%
\footnote{This statement only holds when considering sufficiently long temporal
sequences of snapshots: the individual simulation snapshots at a given
instant in time are not in hydrostatic equilibrium.%
} To elucidate this further, we analyze the horizontal and time-average
of the momentum equation 
\begin{equation}
\left\langle \partial_{t}\rho\vec{v}\right\rangle =\left\langle -\vec{\nabla}\cdot(\rho\vec{v}\vec{v}+\underline{\underline{\tau}})\right\rangle -\left\langle \vec{\nabla}p_{\mathrm{th}}\right\rangle +\left\langle \rho\vec{g}\right\rangle \label{eq:momentum_averaged}
\end{equation}
with $p_{\mathrm{th}}$ being the thermodynamic pressure, $\vec{v}$
the velocity field, and $\underline{\underline{\tau}}$ the viscosity
stress tensor. Due to the averaging, the only remaining spatial dependence
is the vertical one. Divergence terms thus reduce to vertical derivatives,
i.e., $\nab\cdot\left\langle X\right\rangle =\partial_{z}\left\langle X\right\rangle $.
The time derivative $\left\langle \partial_{t}\rho\vec{v}\right\rangle $
vanishes on time average as our model atmospheres are relaxed, hence
quasi-stationary. The inertial term reduces to turbulent pressure
$p_{\mathrm{turb}}=\rho v_{z}^{2}$, so we obtain $\left\langle \vec{\nabla}\cdot\left(\rho\vec{v}\vec{v}\right)\right\rangle =\partial_{z}\left\langle p_{\mathrm{turb}}\right\rangle $.
The divergence of the viscous stress tensor, $\vec{\nabla}\cdot\underline{\underline{\tau}}$,
vanishes on average. The last two terms yield $\partial_{z}\left\langle p_{\mathrm{th}}\right\rangle $
and $\left\langle \rho g\right\rangle $, and we retrieve the equation
for hydrostatic equilibrium with 
\begin{equation}
\partial_{z}\left(\left\langle p_{\mathrm{turb}}\right\rangle +\left\langle p_{\mathrm{th}}\right\rangle \right)=-\left\langle \rho\right\rangle g.\label{eq:hse}
\end{equation}
In Fig. \ref{fig:hse} we show the hydrostatic equilibrium in the
form of $\rho gdz/dp_{\mathrm{tot}}=1$ for the \emph{temporal} and
\emph{geometrical} averaged $\havz$ stratifications, which are very
close to hydrostatic equilibrium. We emphasize that the hydrostatic
equilibrium is only fulfilled by considering the \textit{total} pressure
$p_{\mathrm{tot}}$, as given in Eq. \ref{eq:hse}, which includes
the non-thermal \emph{turbulent pressure} that occupies a significant
fraction of $p_{\mathrm{tot}}$ at the top and in the SAR (see Fig.
21 in Paper I).

Furthermore, one can find in Fig. \ref{fig:hse} that the averages
on a new reference depth scales feature distinctive deviations from
hydrostatic equilibrium (see $\havr$ and $\havcm$). The transformation
of to a new reference depth scale maps all three components of Eq.
\ref{eq:hse} -- geometrical depth $z$, density $\rho$, and total
pressure $p_{\mathrm{tot}}$ -- away from its hydrostatic equilibrium
state. Also, the geometrical depth $z$ loses its strict physical
meaning through such a transformation as a mean value. The mean stratifications
on constant Rosseland optical depth $\havr$ deviate slightly at the
top and significantly in the SAR from the hydrostatic equilibrium
($\havf$ is very similar). The largest departures can be found in
the SAR. Furthermore, the amplitude of the discrepancy from hydrostatic
equilibrium increases for higher $\teff$ and lower $\logg$.

\subsection{Deviations from the EOS\label{app:Deviations-from-EOS}}

\begin{figure}
\includegraphics[width=88mm]{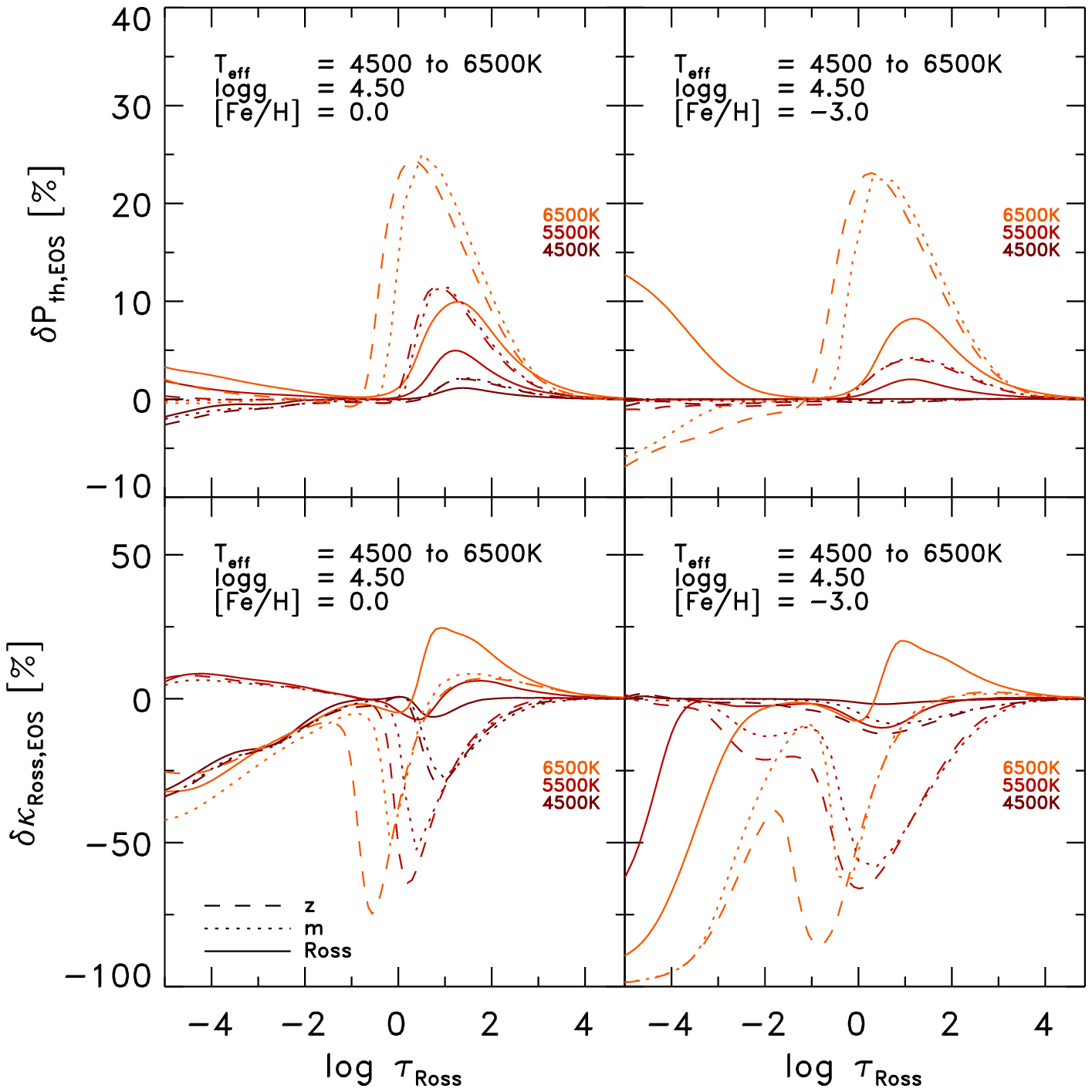}

\caption{Deviations between the spatially and temporally averaged pressure
(\emph{top}) and opacity (\emph{bottom}) and the values derived from
the EOS, i.e. $X\left(\left\langle \rho\right\rangle ,\left\langle \varepsilon\right\rangle \right)$,
vs. optical depth. \emph{Dashed lines}: $\havz$ averages; \emph{dotted
lines}: $\havcm$; \emph{solid lines}: $\havr$.}

\label{fig:kapr_eos} 
\end{figure}

In 3D RHD simulations, the thermodynamic state of a simulation is
self-consistently determined by the EOS. This means in particular
that any thermodynamic variable depends on only two independent variables
(namely the gas density $\rho$ and the internal energy $\varepsilon$)
in a well-defined way. However, the internal self-consistency is broken
by reductions like temporal or spatial averaging.

This can be easily understood by investigating the behavior of a function
$f(X)$ on a 3D cube of quantity $X$. For small fluctuations $X'=X-\left<X\right>$
around the horizontal average at a given depth in the model atmosphere,
a Taylor-expansion of $f$ up to second order yields 
\begin{eqnarray}
f\left(X\right) & = & f\left(\left\langle X\right\rangle +X'\right)\label{eq:mom1}\\
 & \approx & f\left(\left\langle X\right\rangle \right)+\left.\frac{df}{dX}\right|_{\left\langle X\right\rangle }X'+\left.\frac{1}{2}\frac{d^{2}f}{dX^{2}}\right|_{\left\langle X\right\rangle }X'^{2}.\label{eq:mom2}
\end{eqnarray}
The horizontal average of this expression evaluates to 
\begin{eqnarray}
\left\langle f\left(X\right)\right\rangle  & \approx & \left\langle f\left(\left\langle X\right\rangle \right)\right\rangle +\left.\frac{df}{dX}\right|_{\left\langle X\right\rangle }\left\langle X'\right\rangle +\left.\frac{1}{2}\frac{d^{2}f}{dX^{2}}\right|_{\left\langle X\right\rangle }\left\langle X'^{2}\right\rangle \label{eq:mom3}\\
 & = & f\left(\left\langle X\right\rangle \right)+\left[\frac{1}{2}\frac{d^{2}f}{dX^{2}}\left\langle X\right\rangle ^{2}\right]\delta X_{\mathrm{rms}}^{2},\label{eq:mom4}
\end{eqnarray}
where the definition of the contrast $\delta X_{\mathrm{rms}}$ was
used in the last equation (see Eq. \ref{eq:contrast} in Sect. \ref{sub:Contrast}).
The linear term in Eq. \ref{eq:mom3} vanishes as $\left\langle X'\right\rangle =0$
by definition. It is immediately clear that $\left\langle f\left(X\right)\right\rangle =f\left(\left\langle X\right\rangle \right)$
holds for linear functions. It is thus the non-linearity of $f$ that
causes a departure of $\left\langle f\left(X\right)\right\rangle $
from $f\left(\left\langle X\right\rangle \right)$, because the departure
scales with the square of the contrast $\delta X_{\mathrm{rms}}$.
The discussion can be easily expanded to functions of two variables
$f(X,Y)$, since they are found in the EOS.

As a consequence, deriving thermodynamic quantities from averaged
independent variables, $\left\langle \rho\right\rangle $ and $\left\langle \varepsilon\right\rangle $,
will lead to inconsistent outcomes. The mean pressure in a given layer
of the 3D cube will deviate from the pressure calculated with the
EOS from mean density and mean internal energy, $\left\langle p_{\mathrm{th}}\right\rangle \not=p_{\mathrm{th}}\left(\left\langle \rho\right\rangle ,\left\langle \varepsilon\right\rangle \right)$.
Therefore, with $\hav$ we face another level of complexity.

To quantify the deviations, we compute the temperature $T$, pressure
$p_{\mathrm{th}}$, opacity $\kapr$, and electron number density
$n_{\mathrm{el}}$ from the EOS by employing the mean independent
variables $\left\langle \rho\right\rangle $ and $\left\langle \varepsilon\right\rangle $.
Then, we determine the relative disagreement as $\delta X_{\mathrm{EOS}}=(\bar{X}_{\mathrm{EOS}}-\bar{X})/\bar{X}.$
In Fig. \ref{fig:kapr_eos}, we display the deviations of thermal
pressure $\delta p_{\mathrm{th}}^{\mathrm{EOS}}$ and opacity $\delta\kappa_{\mathrm{Ross}}^{\mathrm{EOS}}$.
As suggested by Eq. \ref{eq:mom4}, we find the maximal deviations
typically below the optical surface in the SAR, where the large fluctuations
take place due to the overturning and to the presence of convective
motions with their highly asymmetric up and downflows. The mean value
thus toddles between the bimodal distribution. Furthermore, we find
a strong variation in the $\delta X_{\mathrm{EOS}}$ with stellar
parameter, which increases for higher $\teff$ and lower $\logg$.
Depending on which reference depth scale is applied, the disagreement
$\delta X_{\mathrm{EOS}}$ are distinct.

This loss of consistency caused by dimensional reduction means that
mean $\hav$ models can never entirely substitute full 3D models,
especially for spectral line formation applications \citep{Uitenbroek:2011p10448}.
The mean stratifications are nothing more than statistically meaningful
representations of stellar atmospheres, while only the complete 3D
data set describes their physical state completely. In 1D model atmospheres,
such internal consistency is maintained at all times, since no spatial
averaging of non-linear variables is involved in the construction
of 1D models.
\end{document}